\pdfoutput=1
\documentclass{article}

\usepackage[preprint]{neurips_2026_vericode}

\usepackage[utf8]{inputenc} 
\usepackage[T1]{fontenc}    
\usepackage{hyperref}       
\usepackage{url}            
\usepackage{array}
\usepackage{booktabs}       
\usepackage{amsfonts}       
\usepackage{nicefrac}       
\usepackage{microtype}      
\usepackage{xcolor}         
\usepackage{listings}
\usepackage{amsmath}
\usepackage{amsthm}
\usepackage{tikz}
\usetikzlibrary{shapes.geometric, arrows.meta, positioning}

\IfFileExists{t1lmtt.fd}{%
  \makeatletter\input{t1lmtt.fd}\makeatother
  \DeclareFontShape{T1}{lmtt}{bx}{n}{<-> ec-lmtk10}{}%
  \DeclareFontShape{T1}{lmtt}{bx}{sl}{<-> ec-lmtko10}{}%
  \DeclareFontShape{T1}{lmtt}{bx}{it}{<-> ec-lmtko10}{}%
}{}
\definecolor{keywordcolor}{rgb}{0.7, 0.1, 0.1}   
\definecolor{tacticcolor}{rgb}{0.0, 0.1, 0.6}    
\definecolor{commentcolor}{rgb}{0.4, 0.4, 0.4}   
\definecolor{symbolcolor}{rgb}{0.0, 0.1, 0.6}    
\definecolor{sortcolor}{rgb}{0.1, 0.5, 0.1}      
\definecolor{attributecolor}{rgb}{0.7, 0.1, 0.1} 

\newcommand{\code}[1]{\lstinline{#1}}

\title{SeQLean: Verifying SQL queries using Lean}

\author{%
  Siddhartha Gadgil\\
  Professor, Department of Mathematics, Indian Institute of Science, Bangalore \& \\
  Chief Scientist, Emergence India Labs, Bangalore\\
  \texttt{siddhartha.gadgil@gmail.com} \\
  \And
  Anirudh Gupta \\
  Indian Institute of Science, Bangalore \\
  \texttt{anirudhgupta@iisc.ac.in} \\
  \And
  Ajay Kumar Nair \\
  Research Scientist, Emergence India Labs, Bangalore\\
  \texttt{ajay.nair@emergence.ai}
}
\makeatletter
\@anonymousfalse
\makeatother

\begin{document}

\maketitle

\begin{abstract}
  We present \emph{SeQLean}, a system that uses formal verification in Lean to increase the accuracy of translation from natural language to SQL. This is based on modelling SQL in Lean via relational algebras. Since we do not have a translation known to be correct, we cannot directly prove correctness. Instead, we use \emph{proof challenges}: tests of internal consistency that are unlikely to be satisfied by an incorrect translation. The simplest proof challenge is based on generating two translations of the same natural language query into SQL, ideally with one being optimized for performance and the other for understandability. If two translations are equivalent, they must give the same answer for every database. This can be checked using proof automation in Lean. Using Lean allows us to have essentially arbitrary properties as proof challenges, and also check these (including equivalence) taking into account known properties of the underlying database system. Even for equivalence queries of an arithmetic nature, we show that Lean can be more powerful than traditional tools due to the availability of Mathlib theorems. A JSON API over HTTP allows the verifier to be integrated into agentic frameworks.
\end{abstract}

\section{Introduction}

Querying databases with SQL is a core part of data
analysis performed by agentic AI systems, and the queries are often obtained
by translation from natural language. If the translation is incorrect, the
resulting analysis becomes suspect. It is therefore important to ensure
that translation to SQL has high accuracy.

We present \emph{SeQLean}, a system that uses formal verification in Lean to
increase the accuracy of translation from natural language
to SQL. Since we do not have a translation known to be correct (otherwise
we would simply use it), we cannot directly prove, or even state, what
correctness means. Instead, we use \emph{proof challenges}: tests of
internal consistency that are unlikely to be satisfied by an incorrect
translation.

\subsection{Proof Challenges}
\label{sec:key-idea}

Concretely, the simplest proof challenge is to ask a large language model (LLM) to generate more than
one translation into SQL. If all the translations are correct, they must
be equivalent: they must give the same answer for every database. This
equivalence can be formulated as a mathematical statement in Lean, and
proof automation (building on Lean's tactics, especially \code{grind}) can attempt to prove it. Failure to prove equivalence,
or, better still, finding a counterexample (for which we use the \code{plausible} tactic), indicates that at least one
of the queries may be incorrect; a counterexample proves that at least
one is incorrect.

Conversely, if two translations are equivalent but incorrect, they must
not only both be wrong, but must be wrong in exactly the same way. This is
unlikely, especially if the prompts encourage the two translations to be
structurally different. SQL queries are generally optimized for
performance and concision. If we generate one translation with a default
prompt and a second with a prompt asking for a query optimized for ease of
understanding, we are likely to obtain different translations. Moreover,
a query optimized for understandability is more likely to be correct than
one optimized for performance. Equivalence of two translations generated
in this fashion is therefore strong evidence for the correctness of both.

We emphasize that proving equivalence does not mean testing the queries on
a particular database. We specify only the database schema -- the names and
data types of the columns -- and prove that the two queries return equal
results for every database with this schema. Although we cannot inspect
all such databases, we can abstract the queries as mathematical functions
and attempt to prove that these functions are equal.

There are several natural refinements which we have implemented. The most obvious is to generate
multiple translations and attempt to prove that all are equivalent. We can
also generate auxiliary queries and prove relations weaker than
equivalence. For example, the desired query's result might be required to
be a subset of the result of an auxiliary query. Since all voters must be
at least 18 years old, if the goal is to retrieve all eligible voters in a
constituency, we can also translate a query that retrieves everyone aged
at least 18 and then prove that the first result is contained in the
second.

Further, the columns of a database are not independent. Some may be
computed from others, while others satisfy known relations or constraints.
Queries may therefore become equivalent, or one result may become a subset
of another, only under these constraints. We implement checking under such constraints.

Existing checkers do accept constraints, but only from a fixed catalogue: keys, foreign keys, and per-tuple \texttt{CHECK} predicates. A \texttt{HYPOTHESIS} in SeQLean is an arbitrary Lean proposition about the tables, so it can also state a functional dependency between two columns, a bijection between two columns' value sets, or an arithmetic invariant such as \code{TOTAL = QTY * PRICE}. Appendix~\ref{sec:trusted-constraints} measures where the catalogue stops: of sixteen constraint-dependent rewrites, six rest on a constraint that no such catalogue can state. This also has a use of its own: given a pool of candidate constraints, SeQLean reports which of them a rewrite genuinely depends on, so a rewrite an optimizer would apply silently becomes a short list of properties to validate. Furthermore, due to the availability of Mathlib theorems, Lean can prove equivalence of queries that are arithmetic in nature, which is beyond the capabilities of traditional SQL equivalence checkers.

Throughout, we treat a query's result, and a database table, as a set of rows: duplicates are not distinguished. This matches how real databases are used far more closely than it might first appear. A base table is usually duplicate-free in practice, since a relational primary key, or the object-relational mapping layer used by most web applications, such as Active Record or Django, enforces uniqueness~\citep{zhang2022blockaid}; and where duplicate rows do appear, they are generally treated as an error to be found and removed rather than as meaningful data~\citep{elmagarmid2007duplicate}. Query equivalence is consequently well studied under set semantics specifically, alongside the bag and mixed semantics that duplicate-sensitive queries require~\citep{cohen2009equivalence}. The choice therefore tracks how relational data is actually maintained, rather than being a convenient but unrealistic assumption.

\paragraph{Proving versus Testing.}

An alternative that does not use formal verification is to randomly
generate a collection of databases and test whether query equivalence, or
other expected properties, holds on each of them. This approach can miss
corner cases, for instance differences between strict and weak inequalities, i.e.,
between $<$ and $\leq$. Further, queries typically have several conditions, and so differences such as those between strict and weak inequalities will be detected only for borderline points that satisfy all the other conditions (for example, only if the simulated database contains an individual born on a specific day in a constituency).

\subsection{Applications}

One application of SeQLean is to detect query injection attacks. An attacker may attempt to inject a malicious query into a legitimate one, for example by appending \texttt{OR 1=1} to a \texttt{WHERE} clause. If the original query and the injected query are equivalent, the injection has no effect. If they are not equivalent for \emph{some} database, the injection has changed the query's meaning, and the system can flag this as a potential attack. The counterexample the search returns is then a verified exploit rather than a heuristic alarm. We discuss our implementation in Appendix~\ref{sec:injection-appendix}.

A second application is auditing constraint-dependent rewrites. Many optimizer rules fire only because the schema declares a constraint, and are wrong on data that violates it. Given a rewrite and a pool of candidate constraints, SeQLean reports which of them the rewrite actually depends on: each required constraint comes with the database that breaks the rewrite without it, and each discarded one with the proof that went through anyway. We discuss more in Appendix~\ref{sec:trusted-app}.

\subsection{Related Work}
\label{sec:related-work}

Several recent systems automatically decide or verify SQL query equivalence, primarily to support query optimization and rewrite-rule verification rather than translation checking. \textbf{VeriEQL}~\citep{he2024verieql} encodes query semantics over symbolic tuples in the theory of integers with uninterpreted functions and decides equivalence, up to a chosen database size, with an SMT solver, supporting bag semantics, NULLs, and integrity constraints. \textbf{QED}~\citep{wang2024qed} instead targets equivalence checking that is unbounded and complete for a defined query fragment, representing queries as \emph{Q-expressions} under bag semantics. \textbf{SQLSolver}~\citep{ding2024sqlsolver} reduces equivalence checking to linear integer arithmetic with unbounded sums, also under bag semantics. \textbf{SPES}~\citep{zhou2022spes} takes a two-stage symbolic approach to the same bag-semantics problem. Earlier, \textbf{Cosette}~\citep{chu2017cosette}, whose denotational semantics \textbf{HoTTSQL}~\citep{chu2017hottsql} is machine-checked in Coq based on K-relations, decides equivalence itself not in Coq but by compiling to \textbf{Rosette}~\citep{torlak2014rosette}, a solver-aided host language, and dispatching to its underlying SMT solver -- so, despite the Coq semantics, it is a bounded, SMT-based checker like VeriEQL, not an unbounded proof assistant like ours. HoTTSQL's Coq formalization is nonetheless the prior work closest in spirit to our own use of a proof assistant.

All of these systems answer a single question -- are two given queries equivalent -- typically to justify a query rewrite or migration. While these can be used for checking equivalence as a proof challenge, the side conditions they admit are drawn from a fixed constraint language --- VeriEQL, the most expressive of them on this axis, supports primary and foreign keys, \texttt{NOT NULL}, auto-increment, and per-tuple \texttt{CHECK} predicates built from comparisons, value lists and boolean connectives, but has no arithmetic inside a \texttt{CHECK} and no two-tuple constraint beyond a key or foreign key. A functional dependency, a bijection between two columns, or a denormalization invariant is therefore outside it, while in SeQLean each is an ordinary Lean proposition (Appendix~\ref{sec:trusted-constraints}). A second difference is the nature of the guarantee: VeriEQL and Cosette are sound only up to a chosen database size bound, being SMT-based (the latter via Rosette), whereas QED, SQLSolver, and SPES, like SeQLean, aim for unbounded guarantees -- though under bag rather than set semantics (Section~\ref{sec:key-idea}). We also use the Lean prover~\citep{demoura2021lean4}, together with its mathematical library Mathlib~\citep{mathlib2020}, to prove that the two queries denote the same set of rows, where the systems above rely instead on SMT solvers, Q-expression normalization, or a Coq semantics whose decision step is delegated to a solver.

\section{Implementation}
\label{sec:implementation}

Lean is a programming language whose foundations also allow it to serve as
a formal proof system. It provides excellent support for syntax extensions
and metaprogramming. We use all of these aspects of Lean to verify
equivalence and other properties of SQL queries. A JSON API over HTTP
allows the verifier to be integrated into agentic frameworks.

\subsection{SQL in Lean}

We sketch how we model databases, prove equivalence and other properties, introduce syntax for SQL queries, and find counterexamples in Lean.

\subsubsection{Modelling SQL} \label{sec:modelling-sql}

The first part of the implementation is a model of SQL queries in Lean. We
define types and functions for \emph{database schemas}, \emph{databases} conforming to a
schema, and \emph{SQL queries} over those databases. Our database model is based
on \emph{relational algebra}. A row is represented by a \texttt{TypedTuple},
parameterized by a database schema, and a database is a finite set of such
rows. SQL queries then become functions on databases.

\begin{lstlisting}
abbrev TypedTuple (colType : Fin n -> Type) := (i : Fin n) -> colType i

@[ext] structure TypedRelation 
    (colType : Fin n -> Type) [forall i, DecidableEq (colType i)] where
  labels : Fin n -> String
  rows   : Finset (TypedTuple colType)
\end{lstlisting}

We also define relational-algebra operators used to represent SQL queries
in Lean --- for example, restriction (rows satisfying a predicate) and set difference:

\noindent
\begin{minipage}[t]{0.49\linewidth}
\begin{lstlisting}[basicstyle=\ttfamily\footnotesize]
def restriction
    (predicate : TypedTuple colType -> Bool) (rel : TypedRelation colType) :
      TypedRelation colType :=
  {labels := rel.labels, rows   := rel.rows.filter (fun t => predicate t)}
\end{lstlisting}
\end{minipage}
\hfill
\begin{minipage}[t]{0.49\linewidth}
\begin{lstlisting}[basicstyle=\ttfamily\footnotesize]
def minus (r1 r2 : TypedRelation colType) :
    TypedRelation colType :=
  {labels := r1.labels, rows   := r1.rows \ r2.rows}
\end{lstlisting}
\end{minipage}

Although this modelling step has analogues in any high-level programming
language, the subsequent steps use features specific to Lean.

\subsubsection{Proving Basic Theorems}
\label{sec:basic-theorems}

We next state and prove the basic properties of queries in relational
algebra, together with many supporting results. These proofs serve two
purposes. First, they check the model itself. A common experience when
working with Lean is that subtle errors in definitions are exposed when a
proof reaches a dead end or a claimed theorem turns out to be false.
Second, these theorems can be reused to prove equivalence of concrete
queries obtained by translation. Two examples are:

\noindent
\begin{minipage}[t]{0.40\linewidth}
\begin{lstlisting}[basicstyle=\ttfamily\footnotesize]
theorem inter_absorb_union
    (r1 r2 : TypedRelation colType) :
    intersection r1 (union r1 r2)
      = r1 := by
  ...
\end{lstlisting}
\end{minipage}
\hfill
\begin{minipage}[t]{0.59\linewidth}
\begin{lstlisting}[basicstyle=\ttfamily\footnotesize]
theorem projection_compose {m p : Nat}
    (indices1 : Fin m -> Fin n)
    (indices2 : Fin p -> Fin m)
    (rel : TypedRelation colType) :
    projection indices2 (projection indices1 rel) =
      projection
        (fun j => indices1 (indices2 j)) rel := by ...\end{lstlisting}
\end{minipage}

\subsubsection{Proof Automation}
\label{sec:proof-automation}

The preceding steps let us state query equivalence and prove it
interactively. Autonomous use in agentic frameworks, however, requires
push-button automation: a single function call should attempt to establish
equivalence. No such function can prove every true query equivalence,
because query equivalence is algorithmically undecidable. Our aim is
instead to succeed on a large fraction of the cases encountered in
practice. Appendix~\ref{sec:automation-appendix} gives the planner and one
route in full, and explains why a library of database identities plus
several cheap, backtracking routes recovers more than any single tactic.

Lean's automation is organized around tactics. In particular, the
\texttt{grind} tactic can combine known theorems with several decision
procedures, including SMT-style reasoning. We prove and annotate a
collection of supporting theorems so that \texttt{grind} can use them
effectively. With a sufficiently rich theorem collection and some further
enhancements, this approach proves equivalence in a large number of cases.

\subsubsection{SQL Syntax in Lean}
\label{sec:sql-syntax}

Lean's syntax can be extended with recursively defined syntax categories.
We introduce categories for SQL queries and their components, allowing SQL
queries to occur directly within Lean syntax. A simplified declaration is:

\begin{lstlisting}
syntax "SELECT " (" DISTINCT ")? sql_cols
  " FROM " sql_from
  (" WHERE " term)?
  (" GROUP " " BY " ident,*
    (" HAVING " term)?)?
  (" ORDER " " BY " sql_order_item,*)?
  (";")? : sql_query
\end{lstlisting}

Here, \texttt{sql\_query} is the syntax category corresponding to SQL
queries. We then implement \emph{parsers}: Lean functions that map SQL-query
syntax to terms in our SQL model.

\subsubsection{Declaring Tables and Data Assumptions}
\label{sec:table-hypothesis}

Section~\ref{sec:modelling-sql} showed how a schema is represented in the model. The \texttt{CREATE TABLE} command lets one be declared directly instead, in SQL-like syntax:

\begin{lstlisting}
CREATE TABLE Orders
  (qty INT, price INT, total INT,
   discount FLOAT, valid BOOL)
\end{lstlisting}

This single declaration elaborates into the column-type function \texttt{Orders\_types}, the (initially empty) relation \texttt{Orders : TypedRelation Orders\_types}, and \texttt{Orders\_schema}, a list of column names and types that the SQL parser consults to resolve column references. Column types follow the model of Section~\ref{sec:modelling-sql}: \texttt{INT} becomes $\mathbb{Z}$, \texttt{STRING} becomes \texttt{String}, \texttt{BOOL} becomes \texttt{Bool}, and \texttt{FLOAT} becomes $\mathbb{Q}$ rather than a floating-point type, so that arithmetic on these columns remains exact and provable. A trailing \texttt{NULL} marks a column nullable, wrapping its type in \texttt{Option}.

As discussed in Section~\ref{sec:key-idea}, the columns of a database are frequently not independent, and two queries may be equivalent only in light of such relationships. The \texttt{HYPOTHESIS} command declares exactly this kind of relation between columns and data assumption, reusing the same parser used for queries. For example,

\begin{lstlisting}
HYPOTHESIS qty_le_total : Orders "qty <= total"
HYPOTHESIS total_calc   : Orders "total = qty * price"
HYPOTHESIS disc_rounded : Orders "discount = ROUND(discount, 2)"
HYPOTHESIS all_valid    : Orders "valid"
\end{lstlisting}

declares four assumptions about every row of \texttt{Orders}: a relation between two columns, an arithmetic identity, a rounding fact, and a boolean flag that always holds. Internally, \texttt{HYPOTHESIS name : Table "pred"} parses \texttt{pred} as the \texttt{WHERE} clause of \texttt{SELECT * FROM Table WHERE pred}, extracts the resulting filter, and defines \texttt{name} as the proposition that this filter holds on every row of \texttt{Table}. A hypothesis is never assumed silently: it appears as an explicit argument of any theorem that relies on it, so introducing it cannot compromise soundness. This lets us prove equivalences such as

\begin{lstlisting}
theorem orders_equiv (t : TableRel Orders_schema) (h1 : qty_le_total t) (h2 : total_calc t)
    (h3 : disc_rounded t) (h4 : all_valid t) : (sql%([Orders_schema])
      "SELECT qty * price AS amount,
              ROUND(discount, 2) AS disc
      FROM Orders WHERE qty <= total AND valid") t
  = (sql%([Orders_schema])
    "SELECT total AS amount, discount AS disc
     FROM Orders") t := by sql_equiv
\end{lstlisting}

Two queries that read entirely different columns and expressions, and that would not be equivalent for an arbitrary database of this schema, are proved equal -- but only for databases satisfying the four stated assumptions. This is exactly the more general notion of equivalence mentioned in Section~\ref{sec:key-idea}. A companion form of \texttt{HYPOTHESIS} declares relations between rows rather than within a row: \texttt{HYPOTHESIS emp\_fd : Emp FUNCDEP a -> b} states that any two rows of \texttt{Emp} agreeing on column \texttt{a} also agree on \texttt{b}. Under this assumption, \texttt{GROUP BY a, b} is proved equal to \texttt{GROUP BY a}, since the two group keys then induce the same partition of the rows. 

Such uncoded relationships are the norm in practice, not the exception: real datasets satisfy many functional dependencies, hidden data properties like \code{ROUND(discount, 2) = discount} which play a crucial role in ensuring query-data consistency, key and foreign-key relationships, and arithmetic or business-rule invariants that are never written into the schema or the query, and discovering them is an entire subfield of data profiling~\citep{abedjan2015profiling,papenbrock2015fd}. Query equivalence is well known to be sensitive to exactly these constraints --- two expressions that differ over arbitrary databases can coincide once integrity constraints are assumed, a fact exploited since the earliest work on semantic query optimization~\citep{king1981quist} and characterized theoretically as equivalence under functional and inclusion dependencies~\citep{aho1979equivalences}. A translation checker therefore misses equivalences that genuinely hold on the data at hand whenever the relevant property falls outside the constraint language it accepts; because \texttt{HYPOTHESIS} makes these relationships first-class, explicit inputs to the proof, SeQLean can establish precisely such context-dependent equivalences, and does so soundly --- every assumption is recorded as an argument of the theorem rather than silently baked in.

\subsubsection{Examples of Equivalence}
\label{sec:working-example}

Consider two database tables, \emph{orders} and \emph{customers}, with the
schemas shown in Table~\ref{tab:schemas}.

\begin{table}[t]
  \caption{Schemas of the \emph{orders} and \emph{customers} tables.}
  \label{tab:schemas}
  \centering
  \small
  \begin{minipage}[t]{0.47\linewidth}
    \centering
    \textbf{orders}\\[2pt]
    \begin{tabular}{ll}
      \hline
      \textbf{Column} & \textbf{Type} \\
      \hline
      \texttt{order\_id} & \texttt{INT} \\
      \texttt{cust\_id}  & \texttt{INT} \\
      \texttt{region}    & \texttt{STRING} \\
      \texttt{amount}    & \texttt{INT} \\
      \texttt{status}    & \texttt{STRING} \\
      \hline
    \end{tabular}
  \end{minipage}
  \hfill
  \begin{minipage}[t]{0.47\linewidth}
    \centering
    \textbf{customers}\\[2pt]
    \begin{tabular}{ll}
      \hline
      \textbf{Column} & \textbf{Type} \\
      \hline
      \texttt{cust\_id} & \texttt{INT} \\
      \texttt{name}     & \texttt{STRING} \\
      \texttt{tier}     & \texttt{STRING} \\
      \texttt{active}   & \texttt{BOOL} \\
      \hline
    \end{tabular}
  \end{minipage}
\end{table}

We prove that the following two queries are equivalent:

\noindent
\begin{minipage}[t]{0.49\linewidth}
\begin{lstlisting}[language=SQL, keywordstyle=\color{blue}\bfseries, stringstyle=\color{purple}, basicstyle=\ttfamily\footnotesize]
SELECT
  customers.tier,
  ROUND(SUM(amount), 2) AS total
FROM orders AS o
JOIN customers
  ON o.cust_id = customers.cust_id
WHERE (o.region = 'US' OR o.region = 'EU')
  AND customers.active
  AND customers.tier IN ('gold', 'platinum')
  AND customers.name LIKE 'A%'
  AND o.status = 'completed'
GROUP BY customers.tier
ORDER BY customers.tier
\end{lstlisting}
\end{minipage}
\hfill
\begin{minipage}[t]{0.49\linewidth}
\begin{lstlisting}[language=SQL, keywordstyle=\color{blue}\bfseries, stringstyle=\color{purple}, basicstyle=\ttfamily\footnotesize]
SELECT
  customers.tier,
  ROUND(SUM(amount), 2) AS total
FROM orders AS o
JOIN customers
  ON o.cust_id = customers.cust_id
WHERE customers.name LIKE 'A%'
  AND o.status = 'completed'
  AND customers.tier IN ('gold', 'platinum')
  AND (o.region = 'EU' OR o.region = 'US')
  AND customers.active
GROUP BY customers.tier
ORDER BY customers.tier
\end{lstlisting}
\end{minipage}

Using the parser and embedded SQL syntax, we state the equivalence in Lean
as follows. The push-button tactic \texttt{sql\_equiv} completes the proof.

\begin{lstlisting}
theorem query_equiv :
  sql%([orders_schema, customers_schema])
    "SELECT customers.tier, ROUND(SUM(amount), 2) AS total FROM orders AS o JOIN customers ON o.cust_id = customers.cust_id WHERE (o.region = 'US' OR o.region = 'EU') AND customers.active AND customers.tier IN ('gold', 'platinum') AND customers.name LIKE 'A%' AND o.status = 'completed' GROUP BY customers.tier ORDER BY customers.tier"
  = sql%([orders_schema, customers_schema])
    "SELECT customers.tier, ROUND(SUM(amount), 2) AS total FROM orders AS o JOIN customers ON o.cust_id = customers.cust_id WHERE customers.name LIKE 'A%' AND o.status = 'completed' AND customers.tier IN ('gold', 'platinum') AND (o.region = 'EU' OR o.region = 'US') AND customers.active GROUP BY customers.tier ORDER BY customers.tier" := by
  sql_equiv
\end{lstlisting}

\subsubsection{Counterexamples with \texttt{Plausible}}
\label{subsec:plausible}

Proof automation fails if a theorem (such as query equivalence) is false, but also if it is true but the automation is not strong enough to prove it. To distinguish these cases (with best effort),
we integrate Lean's \texttt{Plausible} library~\citep{plausible}, which helps us to search for counterexamples. We call our adaptation of \code{plausible}, \code{plausible\_sql}. This tactic (as with \code{plausible}) requires two things: a way to sample and shrink a random database, and \texttt{Decidable}/\code{Testable} instances so query equivalence can be evaluated and a failure reported with evidence. We supply both by specifying how a row is sampled as a plain proxy record (interpreted directly into a \code{Finset}, so a sampled database is always a set) and equivalence is made decidable, so a hit is never a guess -- \texttt{Plausible} can only report failure by producing an actual proof that the property fails.

Given a pair of queries, \code{plausible\_sql} searches for a counterexample, i.e., a database on which the queries differ. Namely, \code{plausible} uses the \code{SampleableExt} instances for the table types to randomly generate many small candidate databases -- concrete rows for every table binder -- rather than a few large ones. For each generated database it evaluates both queries on it (the queries are computable or decidable) and checks whether the results are equal. If it encounters a database where the two sides return different results, that database gives a counterexample. The tactic then shrinks the counterexample to a minimal witness and reports it.

When a counterexample is supplied by hand or generated by an LLM, \code{plausible\_sql} checks that it is correct. The candidate counterexample is passed as JSON, which is parsed to build an appropriate statement which is then checked by the \code{decide} tactic.

\vspace{0.05cm}
\begin{minipage}[t]{0.48\linewidth}
\hspace{0.6cm}\textbf{search} (no argument)
\begin{lstlisting}[basicstyle=\ttfamily\footnotesize]

theorem find_counterexample_eg
    (t0 : TableRel EMP_schema)
    (t1 : TableRel DEPT_schema) :
  (sql%([EMP_schema, DEPT_schema])
    "SELECT X.NAME AS XN FROM EMP
     AS X WHERE EXISTS(SELECT *
     FROM DEPT AS Y WHERE Y.LOC=3
     AND X.EMP=Y.MGR AND
     X.DEPT=Y.DEPT)"
  ) t0 t1
  ~= (sql%([EMP_schema, DEPT_schema])
    "SELECT X.NAME AS XN FROM
     EMP AS X") t0 t1
  := by plausible_sql
/- infoview:
these queries are not
equivalent -- plausible found
a database where they differ:
Found a counter-example!
t0 := [(0, 0, 0)]
t1 := []
first query: 0 row(s),
second: 1; 0 only in the
first, 1 only in the second
(4 shrinks) -/
\end{lstlisting}
\end{minipage}
\hfill
\begin{minipage}[t]{0.48\linewidth}
\hspace{0.2cm}\textbf{check} (given a witness)
\begin{lstlisting}[basicstyle=\ttfamily\footnotesize]

theorem not_equivalent_eg :
  Not (forall
    (t0 : TableRel EMP_schema)
    (t1 : TableRel DEPT_schema),
  (sql%([EMP_schema, DEPT_schema])
    "SELECT X.NAME AS XN FROM EMP
     AS X WHERE EXISTS(SELECT *
     FROM DEPT AS Y WHERE Y.LOC=3
     AND X.EMP=Y.MGR AND
     X.DEPT=Y.DEPT)"
  ) t0 t1
  ~= (sql%([EMP_schema, DEPT_schema])
    "SELECT X.NAME AS XN FROM
     EMP AS X") t0 t1)
  := by plausible_sql "[[[0,0,0]],[]]"
/- infoview:
plausible_sql: found a
counter-example -- the
two queries differ on the
supplied database
[[[0,0,0]],[]] -/
\end{lstlisting}
\end{minipage}

\subsection{An LLM Based Proof Search: \texttt{sql\_equiv\_llm}}
\label{subsec:llm}

As the deterministic tactic \code{sql_equiv} may not succeed, we introduce the \code{sql\_equiv\_llm} tactic which uses a large language model to generate proofs. The idea is to ask the LLM to judge whether a pair of queries is equivalent or non-equivalent and to provide a Lean proof or counterexample accordingly. We verify the proposed proof or counterexample in Lean before accepting it.

Before calling the LLM, the tactic first checks whether \code{sql\_equiv} suffices to find a proof or produce a counterexample. If not, we enter a short refinement loop with an LLM (the provider --- Gemini, Claude, or OpenAI --- is configurable) that proposes a Lean tactic block. Crucially, the model is never trusted: every proposal is executed by Lean and kept only if it actually closes the goal in case of a proof; if the model instead judges the queries unequal and returns a witness database, that witness is checked by \code{plausible\_sql} (via the \code{decide} tactic). If elaboration fails, we requery the LLM with the errors and the original context, asking it to try again a few more times so it can correct its mistakes. This ensures that the LLM can extend coverage but never compromise soundness. The prompt sent on every round carries the goal, the table schemas and hypotheses, a goal-relevant slice of the library chosen by premise selection, and --- on a retry --- the model's previous script together with the errors it produced. See Appendix~\ref{sec:prompt-appendix} for more details.

We see that using LLMs genuinely extends coverage (Table~\ref{tab:corpora}) beyond the deterministic approach: measured against the full deterministic tactic, they close \textbf{2} additional Calcite and \textbf{1} additional Literature pairs, and add machine-checked counterexamples on \textbf{7} further Literature pairs.

\subsection{Interfaces}
\label{sec:interfaces}

The components above form a Lean library that can be used interactively by
adding it as a dependency or by opening the project directly. Integration
with agentic frameworks, however, requires access from outside Lean.
Because starting Lean and loading its mathematical library, Mathlib, has a
significant cost, it is important to maintain a running Lean process with
which agents can interact.

Lean itself is implemented in Lean, and the same APIs used by its
interpreter to execute tactics and perform other operations are available
programmatically. Using these APIs, we have implemented a Lean executable
that runs in a loop. It repeatedly accepts JSON containing a database
schema and a collection of queries, then returns JSON reporting the result
of the equivalence proof attempt together with diagnostic information. The
executable is wrapped in a containerized HTTP server, enabling convenient
integration into agentic workflows.

\section{Results}
\label{sec:results}

We evaluated our system on two externally labelled corpora, \emph{Calcite} and \emph{Literature}, under the protocol of Appendix~\ref{sec:methodology}: \emph{Proved} counts \texttt{PROVED} files and \emph{Refuted} counts kernel-verified \texttt{plausible\_sql} witnesses, with the LLM tier credited only for pairs the deterministic tier leaves open. Table~\ref{tab:corpora} gives the ground-truth composition of each corpus alongside how many of each mechanism resolves. Both corpora consist of small, hand-written rewrite-rule pairs; Appendix~\ref{sec:crossskill-appendix} complements them with warehouse-scale queries we derived from Spider2, on which the same tactic proves equivalences for queries reading as many as thirty tables.

\begin{table}[t]
\caption{Ground-truth composition of the two evaluation corpora, how many each mechanism resolves, and the net proved by the whole system. \texttt{sql\_equiv} is the standalone deterministic prover (\emph{Proved}) with \texttt{plausible\_sql} search (\emph{Refuted}); \emph{+LLM only} counts pairs the LLM tier additionally closes beyond step~0, each machine-checked by Lean and independently recompilable without the LLM. \emph{Net} is the total resolved --- proved or refuted --- by \texttt{sql\_equiv} and the LLM together, over all pairs; no pair is both, so the four columns add up to it. Budgets and outcome definitions are those of Section~\ref{sec:methodology}; ground truth follows VeriEQL's evaluation~\citep{he2024verieql,he2024verieqlartifact}.}

\label{tab:corpora}
\centering
\small
\setlength{\tabcolsep}{5pt}
\begin{tabular}{lrrr rr rr r}
\toprule
 & \multicolumn{3}{c}{Ground truth} & \multicolumn{2}{c}{\texttt{sql\_equiv}} & \multicolumn{2}{c}{+ LLM only} & Net \\
\cmidrule(lr){2-4}\cmidrule(lr){5-6}\cmidrule(lr){7-8}\cmidrule(lr){9-9}
Corpus & Pairs & Eq. & Non-eq. & Proved & Refuted & Proved & Refuted & Resolved (\%) \\
\midrule
Calcite    & 397 & 393 & 4  & 110 & 0 & 2 & 0 & 112/397 (28\%) \\
Literature & 64  & 35  & 29 & 26  & 8 & 1 & 7 & 42/64 (66\%) \\
\bottomrule
\end{tabular}
\end{table}

The deterministic tactic proves 110 of the 393 equivalent Calcite pairs and 26 of the 35 equivalent Literature pairs, and finds counterexamples for 8 Literature pairs. We report \emph{no} Calcite refutations, although the search returns 15: none is one of the corpus's 4 non-equivalent pairs, so all 15 are artifacts of our own modelling and we discard the set entire (Appendix~\ref{sec:status-appendix}). What distinguishes the approach from a solver is that these numbers are not a ceiling: the bound is the strength of the proof search, not the expressiveness of an encoding, so any pair that is genuinely equivalent can in principle be proved once the right proof is found. That is what the LLM tier exploits --- prompted only on the pairs the deterministic tier leaves open, it closes 2 further Calcite and 1 further Literature pair and refutes 7 more on Literature, each one kernel-checked. The marginal contribution is modest against the full tactic, but it is strictly additive: the LLM never overturns a deterministic verdict, and everything it produces recompiles without it. Finding the right proof automatically remains the hard part, and it is where the remaining gap lies.

\subsection{Where Lean Wins: Unbounded, Arithmetic Equivalence}
\label{sec:arithmetic-win}

By using Lean, we can make use of its rich mathematical library, Mathlib, to prove equivalences that are out of reach for bounded SMT solvers. Lean's Mathlib contributes tens of thousands of proved theorems about rational and real arithmetic, finite sums, inequalities, and algebra, and our tactics -- mainly \code{grind} -- can invoke them directly. This makes equivalences of an \emph{arithmetic} or \emph{nonlinear} nature, exactly the cases where bounded SMT encodings struggle, provable in a single push-button step. Crucially, the result is \emph{unbounded}: the proof holds for every database of any size, not merely up to a fixed row count.

As an example, consider a filter that keeps each group whose second moment dominates its squared mean:

\begin{lstlisting}[language=SQL, basicstyle=\ttfamily\footnotesize]
-- this can be proved by sql_equiv
-- Q1:  SELECT g FROM obs GROUP BY g HAVING AVG(x*x) >= AVG(x)*AVG(x)
-- Q2:  SELECT g FROM obs GROUP BY g        (the HAVING never removes a group)
\end{lstlisting}

The filter is a no-op (equivalent to the identity) because $\mathrm{AVG}(x^2) \ge \mathrm{AVG}(x)^2$ holds for every group -- it is exactly the statement that variance is non-negative. Our tactic \code{sql\_equiv} discharges it instantly and for all databases: the goal reduces to $\sum_i (x_i-\mu)^2 \ge 0$ over the group, which is \code{Finset.sum\_nonneg} applied to \code{sq\_nonneg} in Mathlib. By contrast, the bounded checker VeriEQL can only confirm this up to a chosen database size (up to 3 on our testbed) and times out as that bound grows. QED reports the above examples and the closely related reduction $\mathrm{AVG}(x{*}x) = \mathrm{SUM}(x{*}x)/\mathrm{COUNT}$ as \emph{NotProvable}, since it lies outside its complete fragment. The gap is not incidental: an unbounded, nonlinear fact about an aggregate over an arbitrarily large relation is precisely what a finite model checker cannot solve and a proof assistant can. 

Furthermore, \emph{any} property expressible in Lean can serve as a proof challenge. Thus SeQLean is not confined to the questions a bespoke solver was built for -- checking any mathematical statement about the queries reduces to writing the goal and letting Lean and Mathlib close it.

\section{Limitations and Future Work}
\label{sec:limitations}

Our system has various limitations which we will address in future work. Firstly, we support a core subset of SQL and some special constructions. However, a production system would need to support a wider range of SQL features. Secondly, our best effort proof automation has to be improved, including integrating with Lean-SMT, to prove or refute in more cases. Thirdly, we have modelled SQL tables as sets -- while this is often the correct model, in other cases Lists or Multisets are more appropriate. We will extend our model to support these cases, and to support the full range of SQL features.

\section{Conclusion} \label{sec:conclusion}

We present SeQLean, which turns SQL query equivalence into a machine-checked proof obligation in Lean. A relation is a \code{Finset} of typed tuples and every operator is a definition, so the algebraic laws the automation uses are theorems rather than assumptions. Each query has one of three outcomes --- a proof holding for every database of the schema, a counterexample database, or no verdict.

Two consequences distinguish this from deciding equivalence with a bespoke solver. Because aggregates carry their arithmetic instead of being uninterpreted symbols, Mathlib settles nonlinear and unbounded identities that bounded SMT encodings cannot reach, and it settles them for databases of any size rather than up to a row bound. And because a side condition is just a Lean proposition, \code{HYPOTHESIS} states integrity constraints that no checker's constraint grammar admits --- functional dependencies, column bijections, arithmetic invariants over a cached column.

An obvious application of equivalence checking is verified optimization of SQL queries. There are other applications as well. An SQL injection results in the intended query and the executed one failing to be equivalent, and a counterexample search returns a concrete exploited database. Another application is to constraint-dependent rewrites -- for these \emph{which} promises they depends on can be traced, with each one justified by the database that breaks the rewrite without it.




\bibliographystyle{plainnat}
\bibliography{references}

\newpage

\appendix

\section{Appendices}
\subsection{Where SeQLean Fits: an Agentic NL-to-SQL Pipeline}
\label{sec:agentic}

\begin{figure}[h]
\centering
\begin{tikzpicture}[
  box/.style={draw, rounded corners, align=center, minimum height=8mm, inner sep=3pt, font=\small},
  hl/.style={draw, rounded corners, align=center, fill=blue!8, very thick, minimum height=10mm, inner sep=3pt, font=\small},
  ok/.style={draw, rounded corners, align=center, fill=green!10, minimum height=8mm, inner sep=3pt, font=\small},
  bad/.style={draw, rounded corners, align=center, fill=red!8, minimum height=8mm, inner sep=3pt, font=\small},
  unk/.style={draw, rounded corners, align=center, fill=black!8, minimum height=8mm, inner sep=3pt, font=\small},
  llmb/.style={draw, rounded corners, align=center, fill=orange!12, minimum height=8mm, inner sep=3pt, font=\small},
  db/.style={draw, cylinder, shape border rotate=90, aspect=0.22, align=center, minimum height=12mm, inner sep=3pt, font=\small},
  ar/.style={-{Latex[length=2mm]}},
  dar/.style={-{Latex[length=2mm]}, densely dotted, thick}
]
\node[box]  (user)   at (0.9, 3.5) {Natural-language\\request};
\node[box]  (agent)  at (4.7, 3.5) {LLM agent\\{\scriptsize two SQL translations}};
\node[db]   (holder) at (10.4,3.5) {Data owners\\{\scriptsize table $+$ \textsc{hypothesis}}};
\node[hl]   (seq)    at (6.4, 1.3) {\textbf{SeQLean}\\{\scriptsize \texttt{sql\_equiv} equivalence check}\\{\scriptsize Lean $+$ Mathlib kernel}};
\node[llmb] (llm)    at (6.4,-0.9) {\texttt{sql\_equiv\_llm}\\{\scriptsize query $+$ relevant context}\\{\scriptsize $+$ Lean errors $\to$ LLM}};
\node[ok]   (proved) at (1.5,-3.1) {\textbf{Done}\\{\scriptsize proof of $Q_A \cong Q_B$}};
\node[unk]  (unknown)at (6.4,-3.1) {\textbf{No verdict}\\{\scriptsize Cannot say anything}};
\node[bad]  (refuted)at (11.0,-3.1) {\textbf{Done}\\{\scriptsize non-equivalence $+$}\\{\scriptsize witness database}};

\draw[ar] (user) -- (agent);
\draw[ar] (agent.south) -- node[left,font=\scriptsize,pos=0.55] {$Q_A,\,Q_B$} ([xshift=-14mm]seq.north);
\draw[ar] (holder.south) -- node[right,font=\scriptsize,pos=0.55] {schema, hypotheses} ([xshift=14mm]seq.north);
\draw[ar] (seq.west) -| node[above,font=\scriptsize,pos=0.25] {proved equivalent} (proved.north);
\draw[ar] (seq.east) -| node[above,font=\scriptsize,pos=0.25] {counterexample} (refuted.north);
\draw[dar] (seq.south) -- node[right,font=\scriptsize] {could not prove} (llm.north);
\draw[dar] (llm.north east) to[out=40,in=-40,looseness=6] node[right,font=\scriptsize,pos=0.5,xshift=1mm] {retry, $\le 5\times$} (llm.east);
\draw[dar] (llm.west) -| ([xshift=6mm]proved.north);
\draw[dar] (llm.south) -- (unknown.north);
\draw[dar] (llm.east) -| ([xshift=-6mm]refuted.north);
\end{tikzpicture}
\caption{SeQLean as the verification step in an agentic NL-to-SQL workflow. The agent emits two translations $Q_A,Q_B$; the data owners contribute the schema and the \textsc{hypothesis} clauses. \texttt{sql\_equiv} either proves the equivalence or returns a counterexample database, both terminal. The dotted path is \texttt{sql\_equiv\_llm}: on a failure it feeds the queries, context and Lean errors back to an LLM for at most five kernel-checked retries, ending in the same three conclusions.}
\label{fig:agentic}
\end{figure}
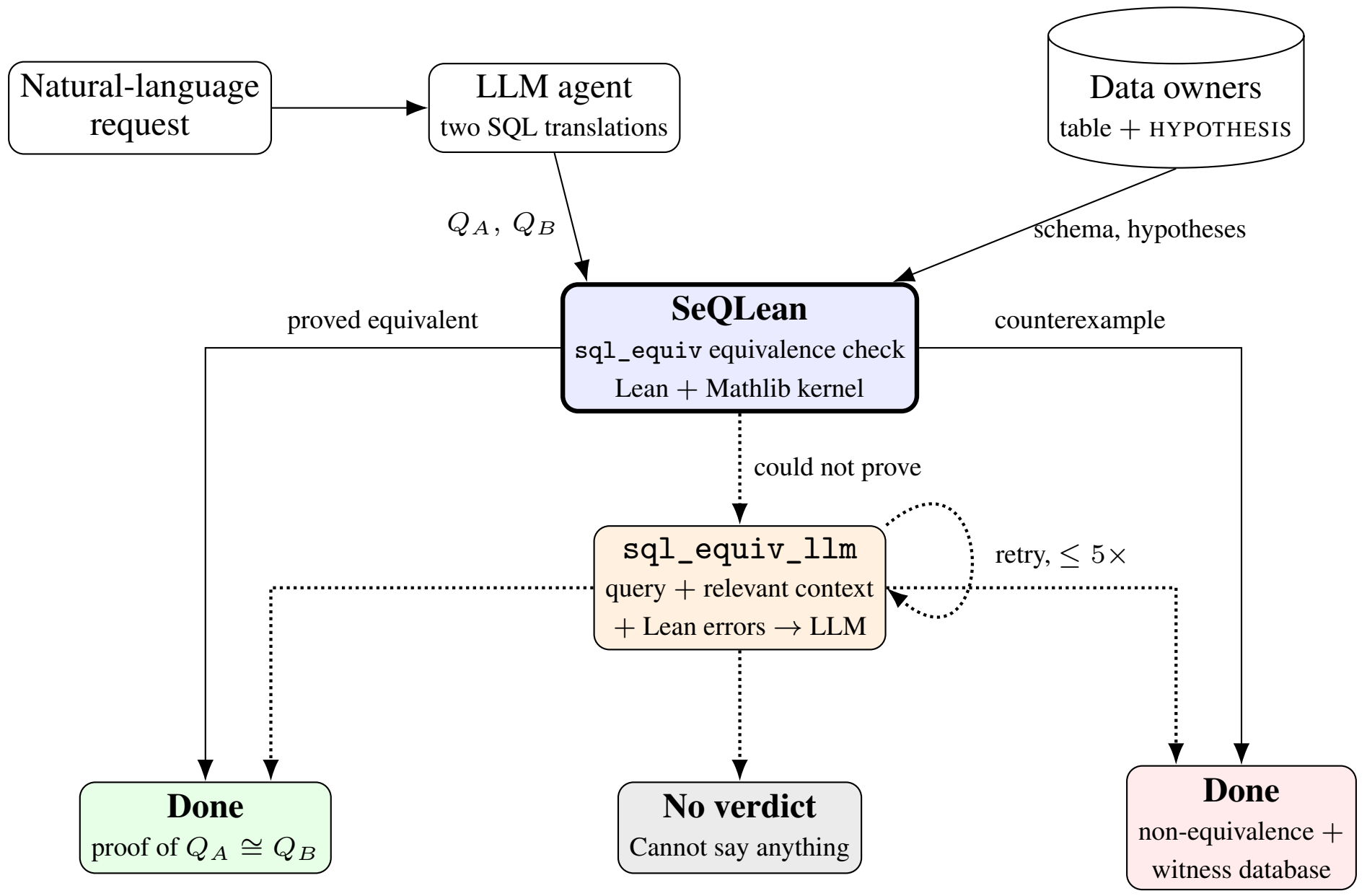

Figure~\ref{fig:agentic} shows the intended deployment. An agent turns a natural-language request into \emph{two} SQL translations (for example one optimized for performance, one for readability). Independently, the owners of the data supply the other half of the obligation: the table schemas and any \textsc{hypothesis} clauses recording integrity constraints and column relationships that hold of their instance but are invisible in the query text (Section~\ref{sec:table-hypothesis}). SeQLean then discharges $Q_A \cong Q_B$ under those hypotheses. A successful proof ends the workflow with a machine-checked certificate over \emph{all} databases of the schema; a counterexample ends it with a concrete witness database and a proof of non-equivalence. Only when \texttt{sql\_equiv} does neither does the dotted path fire: \texttt{sql\_equiv\_llm} hands the two queries, the relevant toolbox context and the failing Lean errors to an LLM, which proposes a repaired proof script that is re-checked by the kernel. That loop runs at most five times and then rejoins the same three conclusions --- proved, refuted, or no verdict --- so nothing the LLM produces can be accepted without a kernel check.

\subsection{Beyond-SMT Equivalences: Moments, Regression, and a Hand Proof}
\label{sec:covar-appendix}

Section~\ref{sec:arithmetic-win} used one variance identity. This appendix collects the rest of that family, over \texttt{obs(g STRING, x FLOAT, y FLOAT)}, and then shows what happens when the automation runs out of lemmas. All of the equivalences below are \emph{unbounded} --- they hold for every database of the schema --- and all are nonlinear in the aggregates, which is the combination bounded SMT encodings cannot produce.

\begin{lstlisting}[language=SQL, basicstyle=\ttfamily\footnotesize]
-- this can be proved by sql_equiv
SELECT g, VAR_POP(x) AS v FROM obs GROUP BY g
   ==   SELECT g, AVG(x*x) - AVG(x)*AVG(x) AS v FROM obs GROUP BY g

-- this can be proved by sql_equiv
SELECT g, COVAR_POP(y, x) AS c FROM obs GROUP BY g
   ==   SELECT g, AVG(x*y) - AVG(x)*AVG(y) AS c FROM obs GROUP BY g

-- this can be proved by sql_equiv
SELECT g, COVAR_POP(x, x) AS v FROM obs GROUP BY g
   ==   SELECT g, VAR_POP(x) AS v FROM obs GROUP BY g

-- this can be proved by sql_equiv
SELECT g, REGR_INTERCEPT(y, x) + REGR_SLOPE(y, x) * AVG(x) AS p FROM obs GROUP BY g
   ==   SELECT g, AVG(y) AS p FROM obs GROUP BY g
\end{lstlisting}

The first two are the moment identities $\mathrm{Var}(x) = E[x^2] - E[x]^2$ and $\mathrm{Cov}(x,y) = E[xy] - E[x]E[y]$; the third says a variable covaries with itself exactly as much as it varies. The fourth is the statement that the least-squares line passes through the group's centroid, and it is the most demanding of the four: the slope is a quotient whose denominator $\mathrm{Var}(x)$ vanishes on a constant group, and the equivalence holds there too. Division is total in Lean, so the degenerate group is not a side condition to be discharged separately --- the same proof covers it. QED reports these outside its provable fragment; VeriEQL has no covariance or regression aggregate to encode, and would in any case answer only up to its row bound.

\paragraph{When the automation runs out: proving by hand.} The push-button tactic is a search over a fixed library of lemmas, so a rewrite resting on a fact the library does not have will not close. Working inside a proof assistant means this is a starting point rather than a verdict. Consider the two-variable analogue of the variance filter:

\begin{lstlisting}[language=SQL, basicstyle=\ttfamily\footnotesize]
SELECT g FROM obs GROUP BY g HAVING AVG(x*y)*AVG(x*y) <= AVG(x*x)*AVG(y*y)
   ==   SELECT g FROM obs GROUP BY g          -- the HAVING removes no group
\end{lstlisting}

The filter is a no-op by the Cauchy--Schwarz inequality. \texttt{sql\_equiv} reduces the goal correctly --- both sides become the same projection, one of them filtered --- and then stops, because nothing in its library says the filter's predicate always holds. What the goal needs is one lemma, and Mathlib already has the underlying inequality over finite sums:

\begin{lstlisting}[language=lean, basicstyle=\ttfamily\scriptsize]
theorem avg_cs (key : TypedTuple colType -> K) (k : K)
    (rel : TypedRelation colType) (f h : TypedTuple colType -> Rat) :
    groupAvgRat key k rel (fun t => f t * h t) * groupAvgRat key k rel (fun t => f t * h t)
      <= groupAvgRat key k rel (fun t => f t * f t)
        * groupAvgRat key k rel (fun t => h t * h t) := by
  simp only [groupAvgRat, groupSumRat, groupCount, div_mul_div_comm]
  have hcs := Finset.sum_mul_sq_le_sq_mul_sq (group key k rel).rows f h
  have hn : (0 : Q) <= ((group key k rel).rows.card : Q) * ((group key k rel).rows.card : Q) :=
    mul_nonneg (Nat.cast_nonneg _) (Nat.cast_nonneg _)
  exact div_le_div_of_nonneg_right (by simpa [sq, pow_two] using hcs) hn
\end{lstlisting}

With that fact instantiated at the query's own two columns, the tactic finishes:

\begin{lstlisting}[language=lean, basicstyle=\ttfamily\scriptsize]
theorem cauchy_filter_is_noop : <the two queries above> := by
  funext r
  have hcs := fun k => avg_cs (fun t => groupKey t) k r (fun t => t 1) (fun t => t 2)
  sql_equiv
\end{lstlisting}

Two things are worth separating here. The proof is not a workaround for a weak model --- the model was right, only the search was short --- and the fact supplied is a general one about group averages, not a fact about this pair of queries, so it extends the library for every later rewrite of the same shape. A bounded solver offers no equivalent move: when its encoding cannot express the property, there is no place to put the missing mathematics. This is also the interface the LLM tier writes into (Section~\ref{subsec:llm}) --- a proposed proof is exactly a term like the one above, executed by Lean and kept only if it closes the goal.

\subsection{How Existing Checkers Work, and Where Each Approach Stops}
\label{sec:mechanisms}

The claims made above about prior systems are claims about their \emph{methods}, so this appendix substantiates them from the methods themselves --- the published algorithms and, where a limitation is visible there, the authors' own artifacts. Table~\ref{tab:mechanisms} summarises. In every case the limitation we point to is one the approach commits to by construction, not an implementation gap that a later release would close.

\begin{table}[h]
\caption{Mechanism of each checker and the point at which its approach stops. The final column names what SeQLean does differently. Sources are the systems' papers and public artifacts; the counts of \texttt{axiom} and \texttt{constant} declarations were taken from UDP's Lean sources.}
\label{tab:mechanisms}
\centering
\small
\begin{tabular}{@{}>{\raggedright\arraybackslash}p{1.6cm}>{\raggedright\arraybackslash}p{3.3cm}>{\raggedright\arraybackslash}p{4.0cm}>{\raggedright\arraybackslash}p{3.3cm}@{}}
\toprule
System & Mechanism & Where the approach stops & SeQLean \\
\midrule
VeriEQL & Symbolic tuples in integers $+$ uninterpreted functions, discharged by SMT & Verdict holds only up to the tuple bound; \texttt{VAR\_POP} and \texttt{STDDEV} are rejected at encoding time & Unbounded; variance is a defined function over \texttt{Finset} sums \\[2pt]
QED & Q-expressions over the semiring $\bar{\mathbb{N}}$; normalize, stabilize, unify, with SMT as oracle & Aggregates are \emph{uninterpreted} higher-order operators --- the largest single failure category in their evaluation & Aggregates carry their arithmetic; \texttt{AVG} is a real $\mathbb{Q}$-valued quotient \\[2pt]
SQLSolver & U-expressions to LIA$^\star$ (LIA with unbounded sums), reduced to LIA via semi-linear sets & Nonlinear terms are deliberately over-approximated, so nonlinear identities are out of reach & Mathlib supplies the nonlinear algebra of finite sums \\[2pt]
SPES & Two-stage symbolic normalization under bag semantics & Semantics lives in the rewrite rules, which must be trusted as axioms & Semantics is a definition; the rules are theorems \\[2pt]
Cosette / HoTTSQL & K-relation semantics in Coq, but equivalence decided by compiling to Rosette and SMT & The Coq semantics does not make the \emph{decision} verified; that step is bounded & The proof \emph{is} the artifact, and is kernel-checked \\[2pt]
UDP & U-semiring normalization, implemented in Lean & The U-semiring is \texttt{constant}s plus $39$ \texttt{axiom}s; results are relative to that axiomatization & $0$ axioms; the laws are proved from definitions \\
\bottomrule
\end{tabular}
\end{table}

\paragraph{Bounded encodings.} VeriEQL~\citep{he2024verieql} encodes both queries over a fixed number of symbolic tuples in the theory of integers with uninterpreted functions and asks an SMT solver whether the two results can differ. This is the most permissive front-end of any system we examined --- it handles NULLs, integrity constraints, sorting and \texttt{CASE} --- and what it buys in exchange is a verdict quantified only over databases of at most the chosen size. Cosette~\citep{chu2017cosette} arrives at the same position from the opposite direction: its semantics HoTTSQL~\citep{chu2017hottsql} is machine-checked in Coq over K-relations, but the equivalence check itself is performed by compiling to Rosette~\citep{torlak2014rosette} and calling an SMT solver, so the mechanized semantics does not extend any guarantee to the decision made about a particular pair. Two consequences follow for both. A bounded verdict is a different proposition from a theorem, and Appendices~\ref{sec:trusted-constraints} and~\ref{sec:crossskill-appendix} exhibit queries whose behaviour changes past the bound; and the encoding grows with the tuple count, so the bound is not a knob that patience can turn.

\paragraph{Unbounded normalization, and the price of an uninterpreted aggregate.} The unbounded systems all model a query as an algebraic object whose equality can be decided by normalization. QED~\citep{wang2024qed} represents queries as Q-expressions over the semiring $\bar{\mathbb{N}} = \mathbb{N} \cup \{\infty\}$ of row multiplicities and runs a normalize--stabilize--unify pipeline with an SMT solver as the oracle for the scalar theory; SQLSolver~\citep{ding2024sqlsolver} translates U-expressions into LIA$^\star$, linear integer arithmetic extended with unbounded summation, and reduces that to plain LIA through semi-linear sets; SPES~\citep{zhou2022spes} normalizes using a set of symbolic rewrite rules. What matters here is how each treats aggregation, since that is where our own gains are concentrated. QED models an aggregate as an \emph{uninterpreted} higher-order operator, and its evaluation reports that this is precisely where it fails: aggregates are the largest single category in its failure breakdown, ahead of custom operators, list semantics and typing. An uninterpreted \texttt{AVG} can be shown equal to itself applied to provably equal bags, but it cannot be unfolded, so \texttt{AVG(x)} $=$ \texttt{SUM(x)/COUNT(x)} is unavailable in principle rather than by omission. SQLSolver meets the same wall from the arithmetic side: its reduction stays sound because nonlinear terms are \emph{over-approximated}, a product of two variables being replaced by a fresh variable --- exactly the step that discards the content of a nonlinear identity. VeriEQL never reaches the question, its encoder rejecting \texttt{VAR\_POP}, \texttt{STDDEV\_POP} and the sample variants outright rather than encoding them.

This is why the variance and covariance identities of Section~\ref{sec:arithmetic-win} and Appendix~\ref{sec:covar-appendix} are not merely hard for the systems above but outside what their methods can express. In SeQLean an aggregate is an ordinary Lean function --- \texttt{AVG} is a genuine $\mathbb{Q}$-valued quotient of a \texttt{Finset} sum by a cardinality, not a symbol --- so such an identity is a statement about finite sums that Mathlib's algebra already knows how to prove.

\paragraph{What is being trusted.} A final difference is visible only in the artifacts. UDP~\citep{chu2018axiomatic}, the ancestor of the semiring line, is itself implemented in Lean, which makes it the sharpest available comparison: it is not the use of a proof assistant that distinguishes this work. The distinction is what the proofs rest on. UDP's SQL layer introduces an uninterpreted \texttt{aggregator} with an uninterpreted denotation, so aggregation is opaque there in exactly the sense it is in QED --- \texttt{AVG(x)} $=$ \texttt{SUM(x)/COUNT(x)} is not a theorem one can reach, because there is nothing to unfold. The same style extends to the algebra beneath it, which is declared rather than constructed: an opaque U-semiring type together with $39$ \texttt{axiom}s asserting the laws it needs, so that the fidelity of those laws to bag semantics is argued on paper rather than checked. SeQLean declares no axioms: relations are \texttt{Finset}s of typed tuples, every operator is a definition, and each algebraic law \texttt{sql\_equiv} uses is a theorem proved from those definitions. The difference is machine-checkable, and we check it --- \texttt{\#print axioms}, applied to our proved equivalences and to the aggregate inequality of Section~\ref{sec:arithmetic-win}, reports only Lean's three standard axioms, \texttt{propext}, \texttt{Classical.choice} and \texttt{Quot.sound}. Where a construct genuinely cannot be modelled under set semantics we leave it unprovable rather than assume it away (Section~\ref{sec:limitations}). What our own side trusts in exchange --- the elaboration path from SQL text to a goal --- is set out in Appendix~\ref{sec:trusted-appendix}.

\paragraph{Where the constraint language stops.} A \texttt{HYPOTHESIS} is an arbitrary Lean proposition, so it is not confined to the constraint grammar a checker fixes in advance; Appendix~\ref{sec:trusted-constraints} measures how far outside that grammar useful rewrites already lie.

\subsection{Multi-Table Equivalences on Warehouse-Scale Queries}
\label{sec:crossskill-appendix}

The Calcite and Literature corpora are rewrite-rule pairs: small, single-table, and written by hand to isolate one optimizer transformation. They are the right benchmark for comparing against optimizer verifiers, but they are not the workload of Section~\ref{sec:agentic}, where a question is answered by a full analytics query. To exercise that setting we derived a handful of equivalence problems ourselves from the Spider2 benchmark's Snowflake warehouses: for a natural-language question we take two independently generated SQL answers that were separately validated as correct, and ask whether they are equivalent as queries. These are the queries analysts actually write --- common table expressions stacked several deep, multi-way joins, conditional aggregation, string concatenation, casts, and in one case thirty daily partition tables unioned together.

Three of them are reproduced below. What makes them tractable for SeQLean is that a proof does not inspect databases at all. Each query elaborates into a term of the relational algebra of Section~\ref{sec:modelling-sql}, and \texttt{sql\_equiv} (Appendix~\ref{sec:automation-appendix}) reduces the two terms with the same congruence and rewriting machinery it uses on a two-line query: a thirty-way \texttt{UNION ALL} is thirty applications of one union lemma, and a chain of common table expressions is a chain of substitutions. The cost therefore tracks the \emph{syntactic} size of the query, not the size or number of the tables it reads, and the resulting theorem holds for every database.

\paragraph{Nested aggregation over composite keys.} \texttt{sf\_local034} asks for the average, across product categories, of the most-used payment method's count, over a Brazilian e-commerce warehouse. Both variants join three tables, count distinct composite keys built by string concatenation, take a per-category maximum, and average the result; they differ in whether the concatenation casts its integer operand explicitly.

\begin{lstlisting}[language=SQL, basicstyle=\ttfamily\scriptsize]
WITH category_payment_counts AS (
  SELECT pr."product_category_name", pay."payment_type",
         COUNT(DISTINCT pay."order_id" || '-' ||
               CAST(pay."payment_sequential" AS VARCHAR)) AS payment_count   -- Q2: no CAST
  FROM OLIST_ORDER_PAYMENTS pay
  JOIN OLIST_ORDER_ITEMS    oi ON pay."order_id"  = oi."order_id"
  JOIN OLIST_PRODUCTS       pr ON oi."product_id" = pr."product_id"
  WHERE NOT pr."product_category_name" IS NULL
  GROUP BY pr."product_category_name", pay."payment_type"),
most_preferred AS (                                        -- Q2: top_payment_per_category
  SELECT "product_category_name", MAX(payment_count) AS top_payment_count
  FROM category_payment_counts GROUP BY "product_category_name")
SELECT AVG(top_payment_count) AS "AVG_TOP_PAYMENT_COUNT" FROM most_preferred
\end{lstlisting}

\texttt{sql\_equiv} proves the pair. The cast is not erased or assumed away: \texttt{CAST(int AS VARCHAR)} is a real function in the model, so the two concatenations are equal because the same function is applied to the same argument, and the outer \texttt{MAX} and \texttt{AVG} then agree by the aggregate congruences. Neither baseline reaches this pair: VeriEQL reports it unsupported and our plan converter cannot express it for QED. The obstacle is shared rather than incidental. \texttt{COUNT(DISTINCT $e$)} over a string-valued expression, feeding a second aggregation over the first one's output, needs a theory of string concatenation that instance-level encodings do not carry and that the decidable fragments of the unbounded checkers are not defined over.

\paragraph{Thirty tables in one query.} \texttt{sf\_ga008} computes, over a Google Analytics event warehouse partitioned into one table per day, the daily page-view total and the average page views per purchasing user during November 2020. The query reads \emph{thirty} tables: a common table expression unions the thirty daily partitions, a second selects the purchasers, the two are joined, and the result is aggregated twice over.

\begin{lstlisting}[language=SQL, basicstyle=\ttfamily\scriptsize]
WITH all_events AS (
  SELECT "EVENT_DATE","EVENT_NAME","USER_PSEUDO_ID" FROM GA4.SAMPLE.EVENTS_20201101
  UNION ALL ... 30 daily partitions ...
  SELECT "EVENT_DATE","EVENT_NAME","USER_PSEUDO_ID" FROM GA4.SAMPLE.EVENTS_20201130),
purchasers AS (SELECT DISTINCT "USER_PSEUDO_ID" FROM all_events WHERE "EVENT_NAME" = 'purchase'),
daily_pv AS (SELECT e."EVENT_DATE", e."USER_PSEUDO_ID", COUNT(*) AS page_view_count
             FROM all_events e JOIN purchasers p ON e."USER_PSEUDO_ID" = p."USER_PSEUDO_ID"
             WHERE e."EVENT_NAME" = 'page_view'
             GROUP BY e."EVENT_DATE", e."USER_PSEUDO_ID")
SELECT "EVENT_DATE", SUM(page_view_count) AS "TOTAL_PAGE_VIEWS",
       ROUND(CAST(AVG(page_view_count) AS DECIMAL), 2) AS "AVG_PAGE_VIEWS_PER_USER"
FROM daily_pv GROUP BY "EVENT_DATE" ORDER BY "EVENT_DATE"
\end{lstlisting}

The two variants differ in the names of their common table expressions and of the intermediate aggregate column. Because \texttt{sql\_equiv} works on the query term rather than on database instances, the thirty partitions cost thirty rewrite steps and nothing more: the whole file elaborates and checks in about ninety seconds on a laptop, and the theorem covers warehouses of any size. Neither baseline returns a verdict: under our configuration the bounded encoding did not finish within its time budget, and our converter does not translate the thirty-way union into a QED plan. The difficulty is structural rather than a matter of engineering, since a method that reasons over symbolic instances pays for thirty tables in the size of the formula it hands the solver.

\paragraph{Conditional aggregation with a large threshold.} \texttt{sf\_bq045} compares rainfall stations in Washington State across two years of NOAA weather data. Two common table expressions each join a yearly observation table to the station table and count rainy days with a conditional aggregate; the two are then joined to each other. The variants differ in whether the \texttt{HAVING} clause names the select alias or repeats the aggregate expression.

\begin{lstlisting}[language=SQL, basicstyle=\ttfamily\scriptsize]
WITH wa_2023 AS (
  SELECT g."stn" AS STATION_ID, g."wban" AS WBAN_ID, s."name" AS STATION_NAME,
         COUNT(CASE WHEN g."prcp" > 0 AND g."prcp" <> 99.99 THEN 1 END) AS RAINY_DAYS_2023
  FROM NOAA.GSOD2023 g JOIN NOAA.STATIONS s ON g."stn" = s."usaf" AND g."wban" = s."wban"
  WHERE s."state" = 'WA' GROUP BY g."stn", g."wban", s."name"
  HAVING RAINY_DAYS_2023 > 150),                -- Q2: HAVING COUNT(CASE WHEN ...) > 150
  wa_2022 AS ( ... the same, over GSOD2022, without the station name ... )
SELECT t23.STATION_ID, t23.STATION_NAME, t23.RAINY_DAYS_2023, t22.RAINY_DAYS_2022
FROM wa_2023 t23 JOIN wa_2022 t22
  ON t23.STATION_ID = t22.STATION_ID AND t23.WBAN_ID = t22.WBAN_ID
WHERE t23.RAINY_DAYS_2023 < t22.RAINY_DAYS_2022 ORDER BY t23.RAINY_DAYS_2023 DESC
\end{lstlisting}

\texttt{sql\_equiv} proves the pair by reducing the \texttt{HAVING} clause to a per-group predicate and observing that the alias and the expression it names denote the same conditional count; the threshold 150 never has to be reasoned about, because no database is ever enumerated. This pair also illustrates what a bound costs on a realistic query. A bounded checker does encode it and reports it equivalent within seconds --- but at the database size we configured, no instance has a station with more than 150 rainy days, so both queries return nothing and the check is confirmed on exactly the instances that cannot exercise it. Raising the bound to where the threshold bites is not a matter of a larger budget: every bounded encoding grows with the number of symbolic tuples, and 150 rows across several joined tables is far outside what an SMT formula of this shape will discharge, whichever system emits it. The unbounded checkers are not limited in this way, but the pair is outside the fragment they decide; QED reports it not provable.

\paragraph{} What these three have in common is not that any particular baseline is weak on them, but that they are the wrong shape for either strategy the field currently uses. A bounded method answers about databases up to a size, and these queries turn on behaviour that only appears well past any tractable bound; an unbounded method answers inside a fixed query fragment, and these queries are built from constructs --- string concatenation under \texttt{COUNT(DISTINCT)}, stacked aggregation, thirty-way unions --- that sit outside the fragments those methods decide. A proof about the query term is subject to neither limit, because it never enumerates a database and never leaves the algebra the query elaborates into. On provenance: the verdicts quoted above are measured only for the two systems we ran (Section~\ref{sec:systems}), those for QED through our own plan converter and therefore a lower bound on what its native front-end achieves. For SQLSolver and SPES we report no measurements and appeal only to the scope their published semantics define.

\subsection{How \texttt{sql\_equiv} Searches for a Proof}
\label{sec:automation-appendix}

\texttt{sql\_equiv} is not a single decision procedure but a small planner over a library of proof routes. This appendix gives its top level, one representative route in full, and the reasoning behind the design.

\paragraph{The planner.} The entry point inspects the goal, then tries the routes in increasing order of cost, each inside its own heartbeat window so that an expensive route cannot starve the ones after it. Rolling back on failure is what makes this safe: a route that does not close the goal leaves no trace.

\begin{lstlisting}[language=ML, basicstyle=\ttfamily\scriptsize]
elab "sql_equiv" : tactic => withMainContext do
  let ty <- instantiateMVars (<- getMainTarget)
  -- classify the goal: constraint-driven / outer-join / aggregate / where- or
  -- projection-congruence / different-base / plain   (traced, for diagnosis)
  trace[sql_equiv] "sql_equiv shape = {shape}"
  -- run a route in a fresh heartbeat window; restore the state unless it closed the goal
  let attempt (bound : Option Nat) (tac : TSyntax `tactic) : TacticM Bool :=
    Core.withCurrHeartbeats <| withTheReader Core.Context (setBudget bound) do
      let s <- saveState
      try
        evalTactic tac
        if (<- getUnsolvedGoals).isEmpty then pure true else do s.restore; pure false
      catch _ => do s.restore; pure false
  let mut closed <- attempt none (<- `(tactic| sql_equiv_safe))   -- normalise + close
  unless closed do closed <- attempt none (<- `(tactic| sql_agg_peel))  -- whole-table aggregates
  unless closed do closed <- attempt none (<- `(tactic| sql_flatten))   -- subquery flattening
  unless closed do Core.withCurrHeartbeats (evalTactic (<- `(tactic| sql_brute_force)))
\end{lstlisting}

\texttt{sql\_equiv\_safe} is the cheap tier: normalise both sides with a congruence loop and close the residual with \texttt{grind} and a handful of algebraic fallbacks. It closes most pairs, and, being membership-free, it can only fail \emph{cleanly} --- which is what lets the LLM tier of Section~\ref{subsec:llm} use it as its step zero and know that a failure is a genuine one. \texttt{sql\_brute\_force} is the last resort and adds the structural route below.

\paragraph{A route in full: reduction to first-order logic over rows.} The most general route replaces the algebra by row membership. \texttt{Finset.ext} turns an equality of relations into $\forall x,\ x \in A \leftrightarrow x \in B$; the \texttt{sql\_mem} lemma set then pushes $\in$ through the algebra --- selection becomes $\wedge$, projection and product become $\exists$, union becomes $\vee$ --- and splits the row equalities that fall out into per-column scalar equalities. What remains mentions only membership in \emph{base} tables: a first-order formula that \texttt{grind} can close. This is the same reduction a bounded checker hands to an SMT solver, except that it is discharged by a proof and therefore quantifies over all databases.

\begin{lstlisting}[language=ML, basicstyle=\ttfamily\scriptsize]
macro "sql_membership" : tactic => `(tactic|
  (try simp only [LeanDatabase.dataEq]
   first | apply Finset.ext
         | (apply TypedRelation.ext (by rfl); apply Finset.ext)
         | skip
   try intro _
   -- first pass: make the appended column lists literal, so the membership lemmas unify at all
   try simp only [List.cons_append, List.nil_append]
   -- second pass: push membership through the algebra and split the row equalities
   try simp only [sql_mem, Finset.mem_image, Finset.mem_filter, Finset.mem_product,
     Finset.mem_union, Finset.mem_inter, Finset.mem_sdiff, Prod.exists,
     decide_eq_true_eq, Bool.and_eq_true, Bool.or_eq_true, Bool.not_eq_true]
   first | grind +locals | sql_constraints | tauto))
\end{lstlisting}

The two-pass \texttt{simp only} is not stylistic. A join's column list reaches the goal as \texttt{l1 ++ l2} inside a \texttt{DecidableEq} instance while the ambient type is already the literal list; until the two agree, no membership lemma unifies --- not even \texttt{Finset.mem\_image} --- and the route silently makes no progress. Diagnosing that class of stall is most of the engineering behind the toolbox.

\paragraph{The other routes.} Each remaining route is a short, guarded script targeting one recurring shape, and each is wrapped in \texttt{done} so that it either closes its goal or backtracks and costs nothing:
\begin{itemize}
\item \texttt{sql\_group\_key} --- two \texttt{GROUP BY}s whose keys induce the same partition. It reduces to the per-group equality and applies the matching aggregate congruence followed by \texttt{group\_congr}, closing the key equivalence by \texttt{simp} alone. Deterministic routes like this are tried before any search, because they are both cheaper and more predictable.
\item \texttt{sql\_hypothesis} and \texttt{sql\_agg\_row} --- equivalences that hold only under a \texttt{HYPOTHESIS}. The first drops a filter the hypothesis makes redundant and finishes the projection row by row; the second descends into the \emph{summed function} of an aggregate, which is the axis a per-row fact travels along, and is the route behind the denormalization rewrites of Appendix~\ref{sec:trusted-constraints}.
\item \texttt{sql\_fk}, \texttt{sql\_funcdep}, \texttt{sql\_bijection} --- integrity constraints. A foreign key \emph{supplies} a parent row for every child row, which is exactly the side condition join elimination needs and the one thing no amount of rewriting can invent; a functional dependency or a bijection shows two keys induce the same partition.
\item \texttt{sql\_flatten} --- subquery flattening, closed by \texttt{aesop} rather than \texttt{grind}. Collapsing the intermediate row a derived table introduces requires choosing existential witnesses, and \texttt{aesop}'s goal-directed search finds those where \texttt{grind}'s congruence closure will not chain them.
\item \texttt{sql\_decide} --- constant relations (\texttt{VALUES} with no table reference) are a closed decidable proposition, so evaluation settles them with no search at all.
\item \texttt{sql\_disprove} --- run first, before any proof attempt, so that a non-equivalent pair is refuted with a witness rather than consuming the whole proving budget (Section~\ref{subsec:plausible}). This is built upon Lean's \texttt{plausible} and is called by \texttt{plausible\_sql} to search for a counterexample.
\end{itemize}.

\paragraph{Why a toolbox rather than one tactic.} Lean's general-purpose automation searches, but it searches over whatever facts are in scope. \texttt{grind} performs congruence closure and E-matching over the lemmas it is given and the hypotheses in context; \texttt{simp} rewrites with the \texttt{@[simp]} set to a normal form; \texttt{aesop} runs a best-first search over a rule set. None of them knows anything about SQL. The database identities that make two queries equal --- that selection commutes with projection, that \texttt{SUM} is additive over a partition, that a distinct count is the cardinality of an image --- are supplied by us, as ordinary proved theorems tagged for \texttt{simp} and \texttt{grind}. Building that library is the actual work; the search is then Lean's.

Two things follow. First, a library beats a monolith because the same lemma is reused by every route: \texttt{group\_congr} is what closes a dropped \texttt{GROUP BY} key, a functional-dependency rewrite, and a bijection rewrite alike, and each new lemma strengthens every route that can reach it. Second, multiple routes beat one strong route because the routes fail in different directions. Normalisation aligns two queries that differ syntactically but not structurally; the membership route handles two queries whose structures differ but whose row sets coincide; the constraint routes handle pairs that are simply false without a hypothesis. A goal that one route cannot even state is often trivial for another, so trying them in cost order --- cheap and deterministic first, search last, each rolled back on failure --- recovers far more than any single tactic, at the price of the failed attempts, which is why every route is budgeted and backtracking. The soundness of the whole arrangement is independent of all of it: whichever route succeeds, the result is a term the Lean kernel checks.

\subsection{What the LLM Tier is Shown}
\label{sec:prompt-appendix}

When \texttt{sql\_equiv} fails cleanly, \texttt{sql\_equiv\_llm} (Section~\ref{subsec:llm}) asks a model for a tactic script. The model is never trusted --- its reply is elaborated by Lean and kept only if it closes the goal --- but what it is shown determines how often it succeeds. The prompt is assembled as a sandwich: framing, then a repo-context block, then the goal and any feedback, then the output instruction restated, since a long input biases attention toward its start and end. Abridged to one or two entries per part, it looks as follows.

\begin{lstlisting}[basicstyle=\ttfamily\scriptsize]
--- FRAMING ---
The goal below is an equality of two elaborated queries, either `A = B` or `A ~= B`
(`dataEq`: rows equal, ignoring output column *labels*). Rows are a `Finset` -- SET
semantics; base tables are assumed duplicate-free.
The repo's own automation `sql_equiv` has ALREADY been run on this goal and did NOT
close it. Do NOT call it again -- write a PRIMITIVE proof.
You may begin with `sql_normalize`, the repo's reduction loop: it collapses BOTH
queries to a small per-row residual. But `sql_normalize` followed by a blind `grind`
is exactly what already failed, so after it use a TARGETED move.

--- RECIPE, SELECTED BY GOAL SHAPE (3 given; one shown) ---
(B) AGGREGATE -- SUM/AVG/COUNT/MIN/MAX, usually with GROUP BY, where the two sides
  differ in a dropped GROUP-BY key or in aggregate arithmetic:
      simp only [TypedRelation.mapByList_rows]
      apply Finset.image_congr; intro x hx
      first | simp only [groupSum_add, groupSum_mul_left, ...]
            | (apply groupSum_congr; apply group_congr; intro _ _; grind +locals)
  COUNT is `Nat` cast to `Int` -- peel with `Nat.cast_inj` / `Int.ofNat.injEq`.

--- PER-CONSTRUCT TIPS (8 given; two shown) ---
- NULL/CASE: `CASE WHEN NOT c IS NULL ...` elaborates to `if !SqlNullable.isNull ...`.
- CONSTRAINT-DEPENDENT: if the theorem's binder has NO hypothesis (`FuncDepEq ...`,
  a UNIQUE fact, a `HYPOTHESIS`), the goal is FALSE -- do not invent a proof; reply
  with the single word `UNPROVABLE`.

--- REPO CONTEXT: theorems first, then operator defs (<= 200 entries; two shown) ---
theorem LeanDatabase.groupSum_congr : (forall t, t in r -> k1 t = k2 t) ->
    groupSum r k1 f = groupSum r k2 f
def LeanDatabase.restriction (r : TypedRelation s) (p : Row s -> Bool) : TypedRelation s

--- GOAL ---
t0 : TableRel [("EMPNO", INT), ("DEPTNO", INT), ("SAL", INT)]
|- sql%(SELECT DEPTNO, SUM(SAL) FROM EMP GROUP BY DEPTNO, JOB) ~= sql%(...)

--- FEEDBACK, ON A RETRY ONLY ---
YOUR PREVIOUS ATTEMPT FAILED:
  unsolved goals: |- groupSum (group t0 ...) = groupSum (group t0 ...)
Return a corrected tactic script.

--- OUTPUT FORMAT (restated) ---
The raw Lean 4 tactic script and NOTHING else: no markdown, no fences, no `by`,
no comments. Never `sorry`, never `sql_equiv`/`sql_membership`.
\end{lstlisting}

The repo-context block is produced by premise selection rather than by dumping the library: a declaration is kept if the goal names it, if its type shares a \emph{discriminative} constant with the goal --- one that is not so ubiquitous across the library as to carry no signal, a threshold computed from the candidate set rather than hard-coded --- or if it is structurally essential to the standard proof shapes. Auto-generated equation and congruence lemmas are filtered out, the uninterpreted scalar functions contribute their theorems but not their definitions, and the block is capped at 200 entries, theorems first. Two details of the framing carry disproportionate weight: telling the model that the deterministic tactic has already failed, which stops it from proposing \texttt{sql\_equiv} back, and giving it \texttt{UNPROVABLE} as an explicit reply, which lets it decline a pair that is genuinely false for want of a constraint instead of fabricating a proof that will be rejected anyway.

\subsection{What is Checked, and What is Trusted}
\label{sec:trusted-appendix}

A proof assistant moves the question of trust rather than eliminating it, so it is worth stating exactly where the boundary sits in SeQLean. On one side of it the guarantee is unusually strong: every operator is a definition and every algebraic law a theorem proved from those definitions, with no axiomatization of the relational algebra in between --- a claim \code{\#print axioms} settles rather than asserts. Once a goal is stated, nothing between it and the kernel is taken on faith.

The same choice fixes the model's scope, and fixes it in the grammar rather than by approximation: what the model observes is the set of rows a query returns, so \code{LIMIT} and \code{OFFSET} are not in the grammar and a query using one fails to elaborate, while \code{ORDER BY} parses and does not change the denotation.

The boundary is the front end: the path from SQL text to that goal is elaboration, not proof, and it is trusted. Three things about it are worth naming, since a reader auditing the artifact will meet them.

\begin{itemize}\itemsep2pt
\item \textbf{Pre-normalization rewrites the query text.} Before parsing, a small set of textual substitutions maps dialect spellings onto the core grammar --- rewriting \code{<> ALL (} to \code{NOT IN (}, deleting \code{IS TRUE} and \code{IS NOT FALSE}, and dropping the default window frame, which is a no-op in SQL. Each is an identity on the fragment the grammar accepts, but they are applied to text and are justified by argument rather than by a theorem.
\item \textbf{One elaboration default.} A scalar subquery projecting a bare column is elaborated as an aggregate over that column, which agrees with SQL exactly when the subquery returns at most one row --- the case such a subquery is written for, and the case SQL itself requires, but not something the elaborator verifies.
\item \textbf{Cross-type row comparison.} When two queries return the same rows under different column labels or types, they are compared through a normalizing erasure of each cell. For the scalar types this is faithful; for array-valued cells it goes through a printed form, so equality there is equality of that rendering.
\end{itemize}

None of these can make the kernel accept a false theorem: they determine \emph{which} proposition gets stated, not whether its proof is valid. That is the right place for the residual risk to sit, and it shrinks as the grammar absorbs the cases the rewrites currently paper over. Ground-truth labels are likewise external, taken verbatim from VeriEQL's evaluation (Appendix~\ref{sec:methodology}), so a disagreement between SeQLean and a label is attributable to SeQLean.

\section{Applications}
\label{sec:applications}

SQL equivalence checking can be used for various applications, including query optimization, SQL Injection detection, verifying trusted databases and more. We present two applications of SeQLean. The first is SQL injection detection: the injected statement is a \emph{different query} from the intended one, so an attack becomes a refuted equivalence and the database the search returns is a verified exploit. The second is constraint-dependent rewriting (Appendix~\ref{sec:trusted-app}): many optimizer rules fire only because the schema declares a constraint, so we prove such a rewrite under an explicit \texttt{HYPOTHESIS} and get a breaking database back the moment it is dropped. Given a pool of candidate constraints, \texttt{minimalHyps} reports which ones the rewrite actually depends on. 

These applications of SeQLean are presented as a proof of concept, and we leave evaluating the performance of SeQLean on these applications to future work. 

\subsection{SQL Injection as Query Inequivalence}
\label{sec:injection-appendix}

SQL injection has been catalogued as CWE-89~\citep{cwe89} for two decades and appears in every edition of the OWASP Top Ten, most recently as A03:2021~\citep{owasp2021}. Its canonical formal account is already stated in the language of this paper. Su and Wassermann~\citep{su2006essence} define a command injection as an input whose substitution changes the \emph{syntactic structure} of the generated command --- the payload stops being a leaf of the parse tree and becomes part of the tree --- and Buehrer et al.~\citep{buehrer2005parse} turn that definition into a runtime check comparing the statement about to execute against the developer's template. Halfond et al.~\citep{halfond2006classification} classify the resulting payload families.

\paragraph{Why an equivalence checker.} ``The server executed something other than what the developer wrote'' is an inequivalence between two queries, and the artefact that demonstrates it is a database on which they disagree. SeQLean answers exactly that, over all databases at once: either the executed statement denotes the same relation as the intended one for every input table, so the payload provably stayed data, or there is a separating database, and that database \emph{is} the exploit.

\paragraph{The vulnerable pattern.} Equivalence checking needs two queries, so injections where the user supplies the whole query are out of scope. The parameter-filling pattern, which produces the overwhelming majority of real vulnerabilities, is not: a developer splices a request parameter into a template, as in
\begin{center}
\texttt{SELECT name FROM users WHERE name = '<INPUT>' AND pass = 'pw123'}
\end{center}
On every input a developer is likely to test --- any input containing no quote --- the naive renderer and a correctly escaping one emit byte-identical statements, which we prove as a theorem. That is why the defect survives testing, and why the recommended fix is structural: a prepared statement fixes the query's structure before any parameter is bound~\citep{thomas2009prepared}.

\subsubsection{Modelling the Two Queries}

String and numeric parameters need separate treatment, because escaping quotes defends the first and does nothing for the second. We model the two queries as follows.

\paragraph{Quoted slots.} For a string parameter, the injection may contain quotations which may change the intended behavior. For example, if the injected code is \code{admin' OR 'a'='a--} then the expression evaluates to True, SQL Comments \code{--} will comment out the password check and we get the data associated with the admin user.

\begin{center}
\footnotesize
\begin{tabular}{ll}
intended & \texttt{SELECT pass FROM users WHERE user='admin' AND pass='pw123'} \\
executed & \texttt{SELECT pass FROM users WHERE user='admin' OR 'a'='a -{}- ' AND pass='pw123'} \\
\end{tabular}
\end{center}

A simple rectification is to double every quote in the input, so that a quote the input carries can no longer close the template's own. For a quoted slot that is enough, and we prove it rather than assert it: \texttt{lex\_renderSafe} says that for \emph{every} input the escaped rendering lexes to the developer's tokens with the input confined to a single string-literal token. The induction is over the escaper, and the side condition \texttt{NotQuoteLed} constrains the template --- the text after the slot must not itself begin with a quote --- not the input.

\begin{lstlisting}[basicstyle=\ttfamily\scriptsize]
def esc : List Char -> List Char                    -- double every quote
  | [] => []
  | c :: cs => if c = '\'' then '\'' :: '\'' :: esc cs else c :: esc cs

def renderSafe (pre u post : List Char) : List Char := renderRaw pre (esc u) post

-- no input escapes its literal
theorem lex_lit_esc (u buf rest : List Char) (h : NotQuoteLed rest) :
    lex (.lit buf) (esc u ++ '\'' :: rest) = .str (buf ++ u) :: lex .top rest := by
  induction u generalizing buf with
  ...

-- the empty-buffer corollary, from the start of a literal
theorem lex_esc (u rest : List Char) (h : NotQuoteLed rest) :
    lex .top ('\'' :: (esc u ++ '\'' :: rest)) = .str u :: lex .top rest :=
  lex_lit_esc u [] rest h

-- quoting is enough: whatever the input is, it is one token
theorem lex_renderSafe (pre u post : List Char) (ts : List Tok) (h : NotQuoteLed post)
    (hpre : forall rest, lex .top (pre ++ rest) = ts ++ lex .top rest) :
    lex .top (renderSafe pre u post) = ts ++ .str u :: lex .top post := by
  rw [renderSafe, renderRaw, hpre, lex_esc _ _ h]
\end{lstlisting}

When the input is spliced in unescaped, \texttt{plausible\_sql} supplies the database on which the two statements differ (Figure~\ref{fig:injection-snippets}, left).

\paragraph{Unquoted slots.} Quoting is sufficient only for a slot that is a string literal. For \texttt{WHERE id = <INPUT>} the payload \texttt{7 OR 1=1} contains no quote, so $\mathit{esc}$ is the identity and the escaped and naive renderings are the same string. The right pair is instead the input taken \emph{as a value} against the input taken \emph{as syntax},
\begin{center}
\begin{tabular}{ll}
intended & \texttt{SELECT pass FROM users WHERE id = 7} \\
executed & \texttt{SELECT pass FROM users WHERE id = 7 OR 1 = 1} \\
\end{tabular}
\end{center}
and the whole password column comes back (Figure~\ref{fig:injection-snippets}, right). An identifier slot such as \texttt{ORDER BY <INPUT>} is worse: quoting is the wrong operation, since wrapping a column name in quotes turns it into a string constant and changes the query on benign input. We prove both statements, so the containment theorem's scope is delimited by theorems rather than by prose.

\begin{figure}[h]
\centering
\begin{minipage}[t]{0.485\textwidth}
\begin{lstlisting}[basicstyle=\ttfamily\scriptsize]
CREATE TABLE users (id INT, uname STRING,
  pass STRING, role STRING)

-- quoted slot: escaped vs. spliced
theorem auth_bypass_tautology :
  Not (forall u : TableRel users_schema,
    (sql%([users_schema])
      "SELECT uname FROM users
       WHERE uname =
         'alice'' OR ''a''=''a'
         AND pass = 'pw'") u
  ~=
    (sql%([users_schema])
      "SELECT uname FROM users
       WHERE uname = 'alice'
         OR 'a' = 'a'
         AND pass = 'pw'") u) := by
  plausible_sql

-- plausible_sql: found a counter-
-- example - the two queries differ
-- on [[[4,"a","pw","a"]]]
\end{lstlisting}
\end{minipage}\hfill
\begin{minipage}[t]{0.485\textwidth}
\begin{lstlisting}[basicstyle=\ttfamily\scriptsize]
CREATE TABLE users (id INT, uname STRING,
  pass STRING, role STRING)

-- unquoted slot: value vs. syntax
theorem numeric_slot_tautology :
  Not (forall u : TableRel users_schema,
    (sql%([users_schema])
      "SELECT pass FROM users
       WHERE id = 7") u
  ~=
    (sql%([users_schema])
      "SELECT pass FROM users
       WHERE id = 7 OR 1 = 1") u) := by
  plausible_sql
-- plausible_sql: found a counter-
-- example - the two queries differ
-- on [[[4,"","",""]]]
\end{lstlisting}
\end{minipage}
\caption{Two attacks as refuted equivalences, taken from the artefact (in ASCII notation for \texttt{$\neg$} and \texttt{$\forall$}; both elaborate exactly as shown). \texttt{CREATE TABLE} declares the schema \texttt{users\_schema} both statements are elaborated against. Each theorem negates an equivalence between the intended and the executed statement; \texttt{plausible\_sql} reports the database it found, and \texttt{decide} has checked the negation on it. Left: the escaped rendering keeps the payload in one literal, the spliced one does not. Right: the slot carries no quote, so escaping is a no-op and the comparison is between the input read as a value and the same input read as syntax.}
\label{fig:injection-snippets}
\end{figure}

\paragraph{Parameters.} Comparing the statement built from one input against the statement built from another conflates two things. The queries for \texttt{name = 'bob'} and \texttt{name = 'alice'} are genuinely inequivalent, and a search returns a separating database --- but nothing is wrong: the difference is data, not behaviour. We therefore model the input as a parameter with its own relation \texttt{param(v)}, a prepared statement expressed inside the query language, so that there is one query family and \texttt{bob} and \texttt{alice} are two values of its parameter. \texttt{sql\_equiv} then proves, over every database and every parameter relation, that the family returns a subset of the authenticated names: a parameter may narrow the answer, never widen it. Benign inputs satisfy this, which is precisely why none of them is an attack; the injected statement violates it. The parameter relation and the safety property are written out below, over a two-column \texttt{users} since the role and id columns play no part here.

\begin{lstlisting}[basicstyle=\ttfamily\scriptsize]
CREATE TABLE users (name STRING, pass STRING)
CREATE TABLE param (v STRING)

-- a parameter may narrow the answer, never widen it: the second
-- statement is the first restricted to the authenticated names
theorem parameter_never_widens
    (t : TableRel users_schema) (p : TableRel param_schema) :
    (sql%([users_schema, param_schema])
      "SELECT users.name FROM users, param
       WHERE users.name = param.v AND users.pass = 'pw123'") t p
  ~= (sql%([users_schema, param_schema])
      "SELECT users.name FROM users, param
       WHERE users.name = param.v AND users.pass = 'pw123'
         AND users.name IN
           (SELECT name FROM users WHERE pass = 'pw123')") t p := by
  sql_equiv
\end{lstlisting}

\subsubsection{Obtaining the counterexamples.}

Each attack is stated as the negation of an equivalence, $\neg\,(\forall t,\ \mathit{intended}(t) \simeq \mathit{executed}(t))$ with $\simeq$ comparing rows ignoring output labels, and discharged by \texttt{plausible\_sql} (Section~\ref{sec:proof-automation}). We do not supply the exploit databases. Random sampling would be useless, since a predicate naming \texttt{'pw123'} is false on every random string; the tactic instead mines the literals out of the elaborated goal together with the pieces \emph{inside} the payload --- \texttt{alice' OR 'a'='a} contributes \texttt{alice} and \texttt{a}, because a payload embeds the values it impersonates --- and draws most generated cells from that pool, so rows collide onto the values the predicates test. Every candidate is settled by \texttt{decide}, so a reported witness carries a kernel-checked proof of the negated proposition: it is not evidence of an exploit but a verified one.

\paragraph{A catalogue.} Table~\ref{tab:injection} lists ten payload classes spanning the families of Halfond et al.~\citep{halfond2006classification}, each stated as two separate statements over the schemas \texttt{users(id INT, uname, pass, role)}, \texttt{secrets(owner, token)} and \texttt{orders(oid INT, cust, total INT)}. The aggregate and \texttt{HAVING} rows show the payload surviving grouping, the number a report is built on inflated to the whole table. The blind-inference row is not an extraction but an oracle: the injected conjunct reads a table the statement has no business touching, and the attacker learns its truth from whether any row comes back. The privilege-escalation row leaves the role guard textually intact and lets \texttt{OR} re-associate around it.

\begin{table}[h]
\caption{Ten injection classes, each refuted by \texttt{plausible\_sql}. The injection column gives the request parameter; intended and executed give the resulting clause text. The counterexample is the database the search discovered and \texttt{decide} verified --- it is the exploit.}
\label{tab:injection}
\centering
\scriptsize
\setlength{\tabcolsep}{3pt}
\begin{tabular}{@{}>{\raggedright\arraybackslash}p{0.10\textwidth}>{\raggedright\arraybackslash}p{0.20\textwidth}>{\raggedright\arraybackslash}p{0.19\textwidth}>{\raggedright\arraybackslash}p{0.19\textwidth}>{\raggedright\arraybackslash}p{0.25\textwidth}@{}}
\toprule
Class & Injection & Intended & Executed & Counterexample \\
\midrule
tautology
  & \texttt{alice' OR 'a'='a}
  & \texttt{name = 'alice'{}' OR '{}'a'{}'='{}'a'}
  & \texttt{name = 'alice' OR 'a'='a'}
  & \texttt{users\{(4,\allowbreak{}"a",\allowbreak{}"pw",\allowbreak{}"a")\}} \\
comment truncation
  & \texttt{alice' -{}-}
  & \texttt{name = 'alice'{}' -{}-' AND pass}
  & \texttt{name = 'alice'}
  & \texttt{users\{(4,\allowbreak{}"alice",\allowbreak{}"pw",\allowbreak{}"pw")\}} \\
union, same table
  & \texttt{x' UNION SELECT pass FROM users -{}-}
  & \texttt{name = 'x'{}' UNION ... -{}-'}
  & \texttt{... UNION SELECT pass}
  & \texttt{users\{(4,\allowbreak{}"x",\allowbreak{}"x",\allowbreak{}"x")\}} \\
union, other table
  & \texttt{x' UNION SELECT token FROM secrets -{}-}
  & \texttt{name = 'x'{}' UNION ... -{}-'}
  & \texttt{... UNION SELECT token}
  & \texttt{+ secrets\{("x",\allowbreak{}"x")\}} \\
numeric slot
  & \texttt{7 OR 1=1}
  & \texttt{id = 7}
  & \texttt{id = 7 OR 1 = 1}
  & \texttt{users\{(4,\allowbreak{}"",\allowbreak{}"",\allowbreak{}"")\}} \\
\texttt{IN}-list
  & \texttt{1, 2, 3}
  & \texttt{id IN (1)}
  & \texttt{id IN (1, 2, 3)}
  & \texttt{users\{(3,\allowbreak{}"",\allowbreak{}"",\allowbreak{}"")\}} \\
aggregate inflation
  & \texttt{alice' OR 1=1 -{}-}
  & \texttt{COUNT(*) ... cust = 'alice'}
  & \texttt{... OR 1 = 1}
  & \texttt{orders\{(3,\allowbreak{}"",\allowbreak{}3)\}} \\
blind inference
  & \texttt{x' AND uname IN (SELECT owner FROM secrets) -{}-}
  & \texttt{uname = 'x'}
  & \texttt{... AND uname IN (SELECT ...)}
  & \texttt{+ secrets\{("",\allowbreak{}"x")\}} \\
privilege escalation
  & \texttt{alice' OR role='admin' -{}-}
  & \texttt{role='user' AND uname='alice'}
  & \texttt{... OR role = 'admin'}
  & \texttt{users\{(4,\allowbreak{}"user",\allowbreak{}"user",\allowbreak{}"admin")\}} \\
\texttt{HAVING} bypass
  & \texttt{100 OR 1=1}
  & \texttt{HAVING SUM(total) > 100}
  & \texttt{... OR 1 = 1}
  & \texttt{orders\{(3,\allowbreak{}"",\allowbreak{}3)\}} \\
\bottomrule
\end{tabular}
\end{table}

\paragraph{Limitations.} A payload confined to \texttt{ORDER BY} changes nothing a set semantics can observe --- a relation is a \texttt{Finset}, which has no order --- so \texttt{sql\_equiv} \emph{proves} the injected query equivalent to the intended one. That is a correct theorem about row sets and a false negative about security, and it is the honest limit of an equivalence-based account: channels outside the relational semantics, including result ordering, error messages and response timing, are not modelled. The containment results are also stated for a concrete template rather than generically over all templates, since the elaborator requires literal query text. The checker's own limits carry over unchanged: a payload built from a construct outside the supported fragment does not elaborate, and the counterexample search is best-effort, so failing to find a separating database is not evidence that the payload is harmless.

\subsection{Trusted Constraints: Which Promises a Rewrite Depends On}
\label{sec:trusted-app}

A \emph{rewrite} is one query replaced by another that should return the same rows --- what an optimizer does whenever it applies a rule~\citep{begoli2018calcite}. Many rules fire only because of a constraint. Seeing a foreign key from \texttt{orders.cust} to \texttt{customers.id}, an optimizer deletes the subquery in \texttt{WHERE cust IN (SELECT id FROM customers)}, on the grounds that it filters nothing out. The problem is that the deletion is correct only if that key is actually enforced: one orphan order, and the rewritten query returns rows the original never did. Nothing in the two queries reveals this, because the licence to rewrite came from the schema rather than from the text --- optimizers have read it that way since semantic query optimization~\citep{king1981quist,aho1979equivalences}. So the input here is the two queries \emph{and} a pool of candidate \texttt{HYPOTHESIS} constraints, and the question is which of them the rewrite actually depends on.

The risk is that a declared constraint and an \emph{enforced} one are different things. Engines already track the distinction --- SQL Server records a foreign key it has not validated (added \texttt{WITH NOCHECK}, or left untrusted after a bulk load) as \texttt{is\_not\_trusted}, and then declines to eliminate a join through it~\citep{sqlserver_trusted}. The gap widens when constraints come from a profiler rather than a DBA: functional-dependency and key discovery return properties that hold of the data \emph{seen so far}~\citep{papenbrock2015fd,abedjan2015profiling}, which is a hypothesis, not a guarantee.

So the question is no longer ``is this rewrite valid?'' but \textbf{``which promises does it actually need, and what breaks if one of them fails?''} We answer both, with a proof in each direction.

\subsubsection{Proving the Rewrite, With and Without the Constraint}

\paragraph{The constraint becomes an antecedent.} \texttt{HYPOTHESIS} (Section~\ref{sec:table-hypothesis}) turns an integrity constraint into a named proposition over the tables, which a theorem then takes as an argument. Every rewrite appears \textbf{twice}, and the two queries are byte-identical between the halves --- the only thing that changes is whether the promise is assumed.

\begin{lstlisting}[basicstyle=\ttfamily\scriptsize]
CREATE TABLE orders    (id INT, cust INT)
CREATE TABLE customers (id INT, name STRING)

HYPOTHESIS orders_fk : orders FOREIGN cust -> customers id

-- TRUSTED: the key is an antecedent, so the semijoin disappears
theorem in_elim (o : TableRel orders_schema) (c : TableRel customers_schema)
    (hfk : orders_fk o c) :
    (sql%([orders_schema, customers_schema])
      "SELECT o.id FROM orders AS o
       WHERE o.cust IN (SELECT c.id FROM customers AS c)") o c
  = (sql%([orders_schema, customers_schema]) "SELECT o.id FROM orders AS o") o c := by
  sql_equiv

-- UNTRUSTED: same two queries, hypothesis deleted, same tactic
example (o : TableRel orders_schema) (c : TableRel customers_schema) :
    ... the two queries above ... := by
  sql_equiv
-- error: these queries are not equivalent - `plausible` found a database
-- where they differ:   o := [(0, 0)]   c := []
\end{lstlisting}

The untrusted half is not a failure message. It is the orphan order that validating the key would have ruled out --- the concrete reason the rewrite needs the promise. Both the trusted and untrusted halves are proved by SeQLean, and the counterexample is verified.

\begin{table}[h]
\caption{Six constraint families, each with the rewrite it licenses and the database that breaks that rewrite once the \texttt{HYPOTHESIS} is removed. Every witness is found by search, not written by hand, and each contradicts its promise directly: the functional dependency says the id determines the department, and the witness is one id with two departments.}
\label{tab:trusted-app}
\centering
\scriptsize
\setlength{\tabcolsep}{4pt}
\begin{tabular}{@{}>{\raggedright\arraybackslash}p{0.28\textwidth}>{\raggedright\arraybackslash}p{0.36\textwidth}>{\raggedright\arraybackslash}p{0.30\textwidth}@{}}
\toprule
\texttt{HYPOTHESIS} & Rewrite it licenses & Breaks without it \\
\midrule
\texttt{orders FOREIGN cust\allowbreak{} -> customers id}
  & drop \texttt{WHERE cust IN\allowbreak{} (SELECT id FROM customers)}
  & \texttt{orders\{(0,\allowbreak{}0)\}}, \texttt{customers\{\}} \\
\texttt{staff "salary >= 0"}
  & drop \texttt{WHERE salary >= 0}
  & \texttt{staff\{(0,\allowbreak{}0,\allowbreak{}-1)\}} \\
\texttt{member UNIQUE id}
  & \texttt{COUNT(DISTINCT id)} $\to$ \texttt{COUNT(*)}
  & \texttt{member\{(-1,\allowbreak{}0,\allowbreak{}0),\allowbreak{} (-1,\allowbreak{}0,\allowbreak{}1)\}} \\
\texttt{assign FUNCDEP id -> dept}
  & \texttt{GROUP BY id, dept} $\to$ \texttt{GROUP BY id}
  & \texttt{assign\{(-1,\allowbreak{}0,\allowbreak{}0),\allowbreak{} (-1,\allowbreak{}1,\allowbreak{}0)\}} \\
\texttt{code BIJECTION id dept}
  & \texttt{COUNT(DISTINCT id)} $\to$ \texttt{COUNT(DISTINCT dept)}
  & \texttt{code\{(0,\allowbreak{}0,\allowbreak{}0),\allowbreak{} (1,\allowbreak{}0,\allowbreak{}0)\}} \\
\texttt{orderlines "total = qty * price"}
  & \texttt{SUM(total)} $\to$ \texttt{SUM(qty * price)}
  & \texttt{orderlines\{(0,\allowbreak{}0,\allowbreak{}0,\allowbreak{}1)\}} \\
\bottomrule
\end{tabular}
\end{table}

Two further constraints are worth naming, because there the rewrite is not syntactic repetition. A range \texttt{CHECK} \texttt{salary >= 0 AND salary <= 500000} absorbs the \emph{different}, weaker filter \texttt{WHERE salary < 1000000}; and a two-value domain \texttt{status = 'open' OR status = 'closed'} turns \texttt{WHERE status <> 'open'} into \texttt{WHERE status = 'closed'}. Neither rewrite holds row by row --- both hold only on the databases the constraint admits.

\paragraph{One promise, many rewrites.} The arithmetic invariant \texttt{total = qty * price} is a cached column a warehouse maintains but never declares, and trusting it once licenses a whole family. \texttt{SUM(total)}, \texttt{AVG(total)} and \texttt{MAX(total)} may each read the cached column instead of recomputing the product; the same holds under a \texttt{WHERE} and per group under \texttt{GROUP BY oid}; \texttt{WHERE total > 100} may read the cached column; and the consistency filter \texttt{WHERE total = qty * price} may be deleted outright, since it removes nothing. \texttt{AVG} is exact here, a $\mathbb{Q}$-valued quotient rather than integer division, so that case is a real average equality. All are proved for every database and any number of rows. Which of these constraints can even be \emph{stated} in the constraint languages automated checkers provide is the subject of the next part --- the functional dependency, the bijection and the arithmetic invariant cannot.

\subsubsection{Which Constraints a Checker Can Even State}
\label{sec:trusted-constraints}

Section~\ref{sec:table-hypothesis} claims that a \texttt{HYPOTHESIS} may state an arbitrary property of the data, not only the constraints a checker's own grammar provides. Every automated equivalence checker fixes such a grammar in advance, because a side condition has to be compiled into the same decision procedure as the queries themselves: the vocabulary is keys, foreign keys, nullability, and at best a per-tuple predicate. To measure where that vocabulary stops we take the most permissive instance of it in the literature --- VeriEQL's constraint grammar~\citep{he2024verieql}, which admits primary and foreign keys, \texttt{NOT NULL}, auto-increment, and \texttt{CHECK} predicates built from comparisons, value lists and boolean connectives, but no arithmetic inside a \texttt{CHECK} and no two-tuple constraint other than a key. QED, SQLSolver and SPES admit strictly less on this axis, so a constraint outside this grammar is outside all of them.

We wrote sixteen rewrites that are false in general and become true under a stated constraint, each proved once and for all databases by \texttt{sql\_equiv} from the corresponding \texttt{HYPOTHESIS}. Table~\ref{tab:trusted} classifies them by the form the constraint takes. Ten are keys, foreign keys or per-tuple range conditions and fall inside the vocabulary above; six do not, and are unstatable as an integrity constraint in any of these systems.

\begin{table}[h]
\caption{Sixteen rewrites that are false in general and true under the stated constraint. \texttt{sql\_equiv} proves every one for all databases from the corresponding \texttt{HYPOTHESIS}, whatever form the constraint takes. The third column records whether that constraint can be \emph{stated at all} as an integrity constraint in the grammars automated checkers provide: a functional dependency, a bijection between two columns' value sets, and an arithmetic invariant cannot, and in the last six rows the side condition exists only as a Lean proposition.}
\label{tab:trusted}
\centering
\small
\begin{tabular}{lllc}
\toprule
Rewrite & Constraint & Statable as a constraint & SeQLean \\
\midrule
\texttt{IN}-elimination            & foreign key            & yes & proved \\
\texttt{EXISTS} anti-join is empty & foreign key            & yes & proved \\
join elimination                   & foreign key            & yes & proved \\
join elimination                   & foreign key $+$ key    & yes & proved \\
redundant \texttt{WHERE}           & \texttt{CHECK} $\ge 0$ & yes & proved \\
absorbed comparison                & range \texttt{CHECK}   & yes & proved \\
negation becomes the other value   & two-value domain       & yes & proved \\
\texttt{COUNT(DISTINCT)} collapse  & unique key             & yes & proved \\
self-join elimination              & unique key             & yes & proved \\
every group is a singleton         & unique key             & yes & proved \\
\texttt{GROUP BY} collapse         & \textbf{functional dependency} & \textbf{no} & proved \\
\texttt{MAX} $=$ \texttt{MIN} per group & \textbf{functional dependency} & \textbf{no} & proved \\
\texttt{COUNT(DISTINCT)} swap      & \textbf{column bijection}      & \textbf{no} & proved \\
\texttt{SUM} over a cached column  & \textbf{arithmetic invariant}  & \textbf{no} & proved \\
filter on a cached column          & \textbf{arithmetic invariant}  & \textbf{no} & proved \\
the invariant filters nothing out  & \textbf{arithmetic invariant}  & \textbf{no} & proved \\
\bottomrule
\end{tabular}
\end{table}

Expressibility is not the only obstacle these rewrites expose. Among the ten whose constraint \emph{is} statable, we ran the pair itself through the baselines of Section~\ref{sec:systems} and two failed for reasons unrelated to the constraint: one query lies outside the supported fragment, and one rewrite is reported non-equivalent because it is judged under bag rather than set semantics --- a divergence shared by every bag-semantics system in this class, QED, SQLSolver and SPES alike, and not a defect of any one of them (Section~\ref{sec:key-idea}).

A third cost is the size bound, and it belongs to the bounded-SMT approach rather than to any single tool. Comparing \texttt{SELECT g FROM t GROUP BY g HAVING COUNT(*) >= 3} against the empty result, a bounded checker configured at two rows per table reports \textbf{EQUIVALENT} --- correctly, since no group of three can exist in two rows --- and non-equivalent at bounds of three and four. A bounded verdict is therefore not a weaker version of a proof but a different statement, and reading it as one is unsound for exactly the queries whose behaviour begins past the bound. The caveat covers Cosette as well, which despite its machine-checked Coq semantics decides equivalence by compiling to Rosette and an SMT solver.

\subsubsection{Which Constraints Must Be Trusted}

\paragraph{\texttt{minimalHyps}: a rewrite becomes a checklist.} The question can also be inverted. The two queries are supplied together with a pool of candidate constraints --- properties the schema \emph{might} have, of the kind a profiler would suggest --- and \texttt{minimalHyps} returns the subset the rewrite actually depends on. It proves the equivalence under all candidates, drops them one at a time and keeps only those whose removal breaks the proof; then, for each survivor, it searches among the databases satisfying \emph{all the other} retained constraints for one where the two queries disagree.

\begin{lstlisting}[basicstyle=\ttfamily\scriptsize]
def denormCandidates := json%
  {"schemas": [{"name": "orderlines", "columns": [
      {"name": "oid",   "type": "Int"}, {"name": "qty",   "type": "Int"},
      {"name": "price", "type": "Int"}, {"name": "total", "type": "Int"}]}],
   "hypotheses": [
     {"name": "total = qty * price", "table": "orderlines",
      "predicate": "total = qty * price"},
     {"name": "qty > 0",      "table": "orderlines", "predicate": "qty > 0"},
     {"name": "oid is a key", "table": "orderlines", "unique": "oid"}],
   "queries": ["SELECT SUM(total) AS s FROM orderlines",
               "SELECT SUM(qty * price) AS s FROM orderlines"]}

#eval do IO.println (<- minimalHypsReportCore denormCandidates)

-- valid - 1 of 3 candidate constraints must be trusted
--   MUST TRUST  total = qty * price
--      without it: orderlines := [(0, 0, 0, 1)]
--   not needed  qty > 0
--   not needed  oid is a key
\end{lstlisting}

Both verdicts are earned, which is the point. \texttt{MUST TRUST} arrives with \texttt{orderlines := [(0,0,0,1)]}: one stale row whose cached \texttt{total} is \texttt{1} while the product is \texttt{0} --- a database satisfying every other candidate on which the rewrite is wrong. \texttt{not needed} is backed by the proof that went through without it. ``Can I ship this rewrite?'' therefore becomes a short list of properties to go and validate, each item justified by a counterexample and each omission by a proof, rather than a yes/no from an engine that has silently assumed whatever the schema happened to declare.

\paragraph{Limitations:} One caveat, about what ``minimal'' means here. The search is a single linear pass, so the drop order decides \emph{which} irredundant set comes back. Suppose a rewrite is licensed either by one functional dependency \texttt{FUNCDEP cust -> region} or, independently, by \texttt{FOREIGN orders.cust -> customers.id} together with \texttt{UNIQUE customers.id}. If the pass tries the dependency first, the proof still goes through --- the other two carry it --- so the dependency is dropped and the report asks the DBA to validate two properties where one would have done. Every constraint in the answer is load-bearing given the others kept, but the answer need not be the smallest such set. Fixing this means searching subsets rather than making one pass, at a full proof search per attempt; for the candidate pools a real schema produces we took the cheap pass.

\section{Evaluation Methodology}
\label{sec:methodology}

We evaluate our tool on two corpora of SQL query pairs. Each pair supplies the \texttt{CREATE TABLE} schema of every table it reads, the two queries themselves, and any integrity constraints that are assumed to hold of the data.

\textbf{Calcite} (397 pairs) and \textbf{Literature} (64 pairs) are the two benchmarks used to evaluate VeriEQL~\citep{he2024verieql}, taken from its public artifact~\citep{he2024verieqlartifact}. Calcite pairs are extracted from Apache Calcite's optimizer rewrite rules~\citep{begoli2018calcite}; Literature pairs are collected from the classical query-equivalence literature~\citep{chu2017cosette,chu2018axiomatic,wang2018speeding}. We adopt VeriEQL's ground-truth labels verbatim rather than relabelling: 393 of the 397 Calcite pairs and 35 of the 64 Literature pairs are equivalent, the remaining 4 and 29 non-equivalent. Using an external ground truth matters, because it means a disagreement between SeQLean and the label is attributable to SeQLean and cannot be explained away.

\subsection{From a query pair to a proof obligation}
\label{sec:goalgen}

Each pair is compiled mechanically --- with no human editing --- into a standalone Lean file. The generator emits the \texttt{CREATE TABLE} declarations of Section~\ref{sec:table-hypothesis}, one universally quantified table variable per schema, the two queries inside the \texttt{sql\%} quotation of Section~\ref{sec:sql-syntax}, and the proof script. Two independently written queries frequently agree on their result rows while disagreeing on output column \emph{labels}, so the obligation is emitted with the data-equivalence relation $\cong$, which compares rows while ignoring labels, rather than with plain equality --- uniformly, for all 397 Calcite and all 64 Literature pairs, so that no pair is counted as a failure over an alias. Plain equality is the stronger statement and is what the hand-written examples elsewhere in the paper use. Constraints shipped with a pair become \texttt{HYPOTHESIS} declarations and appear as explicit antecedents of the generated theorem, so a proof that uses one records that fact in its statement.

Every generated file ends in \texttt{:= by first | sql\_equiv | sorry}, and carries \texttt{maxHeartbeats 1000000} and \texttt{maxRecDepth 1000000}. The \texttt{sorry} fallback is what makes the census self-classifying: it separates a clean tactic failure from a hard error, as described next.

\subsection{Outcome classification}
\label{sec:classification}

Each generated file is compiled by its own \texttt{lake env lean} process and classified into exactly one outcome from the process exit code, Lean's own diagnostics, and the source position those diagnostics carry. Two things distinguish the outcomes: \emph{how far} a pair got, and \emph{why} it stopped. UNSUPPORTED SYNTAX never reached a goal at all. The remaining three all reached one and differ in why the search ended --- a timeout means the search was \emph{cut off} while still running, so no verdict was reached and more budget might yet produce one; UNPROVED means the search \emph{ran to completion on its own} and came back empty, so more budget would change nothing.

\begin{itemize}\itemsep2pt
\item \textbf{PROVED} --- exit code $0$ and no \texttt{sorry} warning: the goal was closed. Throughout the paper \textbf{RESOLVED} means PROVED \emph{or} REFUTED --- a pair decided in either direction --- and is never used as a synonym for PROVED.
\item \textbf{UNPROVED} --- exit code $0$ but Lean reports \emph{declaration uses \texttt{sorry}}: the pair elaborated into the model, and \texttt{sql\_equiv} ran every strategy it has, terminated of its own accord, and closed nothing. Because the tactic finished rather than being interrupted, raising the budget cannot turn this into a proof; only a stronger tactic can. This is the one category that is a genuine limit on proving power.
\item \textbf{UNSUPPORTED SYNTAX} --- nonzero exit code with an elaboration diagnostic: the pair never became a goal, because a construct in it is outside the grammar or has no elaboration. This bounds the \emph{front end}, not the prover.
\item \textbf{ELAB-TIMEOUT} --- the budget was exhausted while turning the SQL text into a goal.
\item \textbf{PROVE-TIMEOUT} --- a goal existed and the tactic was still searching when the budget ran out, so it was cut off with no verdict either way; unlike UNPROVED, a larger budget can move a pair out of this category. Two mechanisms produce it: the wall-clock cap on the process, and Lean's deterministic heartbeat limit. The latter needs care, because a deterministic timeout is uncatchable, escapes the \texttt{first} combinator and exits nonzero exactly as an elaboration error does; classifying by exit code alone would file it under UNSUPPORTED SYNTAX. We separate the two by the diagnostic and its source position.
\end{itemize}

Refutations are counted separately and never inferred from a proof failure: a pair is \emph{refuted} only when \texttt{plausible\_sql} produces a concrete database and Lean's kernel accepts a \texttt{decide} proof that the two queries differ on it (Section~\ref{subsec:plausible}). This asymmetry is deliberate --- failure to prove is not evidence of non-equivalence, so the two directions are measured by two different mechanisms and reported in two different columns.

\subsection{Full outcome breakdown}
\label{sec:status-appendix}

Table~\ref{tab:status} partitions every pair of each corpus by the strongest verdict any mechanism reached. The point of reporting it is that the outcomes fail at three different stages, and conflating them flatters or damns the wrong component. UNSUPPORTED SYNTAX means the pair never became a goal, so it bounds the front end; widening the parser converts these directly into goals. A timeout means the pair was cut off with the search still in progress, so it carries no verdict at all; we record separately whether the budget ran out before a goal existed (ELAB-TIMEOUT) or after (PROVE-TIMEOUT). UNPROVED is the opposite situation and should not be read as a longer timeout: the tactic was never interrupted, it exhausted its strategies and returned nothing, neither proof nor counterexample. The practical difference is what would move a pair out of the category --- a bigger budget for PROVE-TIMEOUT, a stronger tactic for UNPROVED --- which is why UNPROVED is the only one of the two that measures proving power. REFUTED is not a failure at all but a decided negative: a database on which the two queries return different rows, checked by the kernel.

Two boundaries here are easy to get wrong, and both cut against us if ignored. A deterministic heartbeat timeout exits nonzero exactly as an elaboration error does, so classifying by exit code alone inflates UNSUPPORTED SYNTAX at the expense of PROVE-TIMEOUT --- on Calcite that misattributes $82$ pairs, roughly a third of what would otherwise be reported as unsupported syntax, and on Literature $12$. We separate them by the diagnostic and its source position, and every such timeout on both corpora is positioned inside the tactic rather than in the query term. ELAB-TIMEOUT is consequently empty here, which is itself worth stating: the front end never runs out of budget, and a pair it cannot handle is rejected in under ten seconds, against a three-hundred-second cap. Likewise a counterexample surfaces as a thrown error inside \texttt{first}, which without care is absorbed into UNPROVED; recovering those is what leaves Literature with no UNPROVED pairs at all.

Read this way the two corpora are limited by different things. On Calcite the dominant obstacle is grammar coverage ($136$), with proof-search budget a close second ($128$); $21$ pairs elaborate, receive a completed search, and defeat it. On Literature almost everything that elaborates is decided one way or the other, and what remains open is coverage and budget rather than proving power.

\paragraph{Why we report no Calcite refutations.} The counterexample search does return $15$ witnesses on Calcite, and we credit none of them. Calcite's pairs are optimizer rewrite rules and all but $4$ are equivalent, so a set of $15$ cannot be mostly genuine; inspecting them confirms it, as they are rules such as pushing a \texttt{CASE} expression through a join into a subquery, or merging a two-level aggregation, which are true equivalences. Each witness is nonetheless a real disagreement between the two \emph{Lean terms}, and the kernel is right to accept it --- the defect is upstream of the proof, in the elaboration of the SQL into those terms, and is exactly the kind of unfaithfulness a corpus of known-equivalent pairs is good at exposing. This is why refutation is reported separately from proving throughout, and why we withdraw the entire set rather than keep whichever part of it might be sound: we cannot currently tell which. The proving direction is unaffected, since a modelling error of this kind makes \texttt{sql\_equiv} fail to prove a true equivalence rather than prove a false one. Literature's refutations are retained --- $29$ of its $64$ pairs are non-equivalent by construction, so the count is consistent with its ground truth --- but they have not been audited pair-by-pair against the published labels, and we flag that as a limitation of the present numbers rather than a verified agreement.

\begin{table}[h]
\caption{Outcome breakdown over all pairs of each corpus. Every pair falls in exactly one of the six rightmost categories, and PROVED, REFUTED and the four open categories sum to the corpus size. RESOLVED means decided in either direction --- PROVED (\texttt{sql\_equiv} or the LLM tier closing the goal) plus REFUTED (a kernel-checked counterexample from \texttt{plausible\_sql} or an LLM-proposed witness) --- and is used in that sense throughout the paper. PROVE-TIMEOUT pools the wall-clock cap with Lean's deterministic heartbeat limit, both positioned inside the tactic; ELAB-TIMEOUT is empty because no pair exhausts its budget before a goal exists. Refutations are credited from both mechanisms, which is why some Literature pairs are REFUTED although the deterministic census alone does not elaborate them. No Calcite refutation is credited; the $15$ the search returns are artifacts of our elaboration rather than genuine inequivalences, and are returned here to the categories they came from.}
\label{tab:status}
\centering
\footnotesize
\setlength{\tabcolsep}{3.5pt}
\begin{tabular}{lrrrrrrrr}
\toprule
& & \multicolumn{3}{c}{RESOLVED} & \multicolumn{4}{c}{Open} \\
\cmidrule(lr){3-5}\cmidrule(lr){6-9}
Corpus & Pairs & PROVED & REFUTED & Total & \shortstack[r]{UNSUPPORTED\\SYNTAX} & \shortstack[r]{PROVE-\\TIMEOUT} & \shortstack[r]{ELAB-\\TIMEOUT} & UNPROVED \\
\midrule
Calcite    & 397 & 112 & 0  & 112 (28\%) & 136 & 128 & 0 & 21 \\
Literature & 64  & 27  & 15 & 42 (66\%)  & 9   & 13  & 0 & 0  \\
\bottomrule
\end{tabular}
\end{table}

\paragraph{Refutation counts, and what integrity constraints do to them.} Our search returns $15$ witnesses on Calcite against a ground truth of $4$ non-equivalent pairs, and we credit none of them. The excess is not a matter of a few doubtful cases: the corpus's four non-equivalent pairs are known, and no witness we return is one of them, so all $15$ are artifacts of our own elaboration rather than genuine inequivalences. We discard the set entire and report $0$, which is why Table~\ref{tab:corpora} carries no Calcite refutation.

The cause is identifiable, and VeriEQL's own evaluation isolates it, since that work runs its counterexample search in two configurations and reports both~\citep{he2024verieqlartifact}. With integrity constraints enabled it returns $4$ counterexamples on Calcite, all $4$ genuine --- exactly the ground truth. With them disabled it returns $23$, of which $1$ is genuine. Our search is in the second configuration: it samples databases directly and does not yet consume the keys and foreign keys a pair declares, so a witness it returns may be a database the declared constraints would have excluded. Our $15$ therefore belongs next to VeriEQL's $23$ rather than next to its $4$ --- and its precision is worse, $0$ genuine rather than $1$. The remedy is identified precisely and on independent evidence: every Calcite pair carries its constraints, enabling them collapses VeriEQL's count to exactly the four true negatives, and constraint-aware sampling is the corresponding change on our side (Section~\ref{sec:limitations}). Until that change lands, a Calcite refutation from our search is not evidence of anything, and we do not present it as such.

On Literature the picture is different, and better, because the corpus is not constraint-dominated in the same way: the deterministic search refutes $8$ pairs and the LLM tier a further $7$, $15$ in all, and these we do credit --- they are consistent with the corpus's $29$ non-equivalent pairs and each is a kernel-checked \texttt{decide} on the witness. VeriEQL nonetheless finds all $29$ with constraints enabled, and we state plainly that refutation is the direction in which a bounded symbolic search has the structural advantage, since it solves for a separating instance rather than sampling for one.

\subsection{Running all comparisons}
\label{sec:systems}

We evaluate SeQLean against two external equivalence checkers, VeriEQL~\citep{he2024verieql} and QED~\citep{wang2024qed}, selected because their guarantees are limited along orthogonal axes. VeriEQL decides any query it can encode, but only over databases up to a fixed size bound; QED's results are independent of database size, but hold only within the SQL fragment for which it is complete. Together they characterise the two principal ways in which an automated checker falls short of a general guarantee. All three systems are run on an identical list of query pairs on a single machine, an AMD Ryzen Threadripper PRO 5975WX (32 cores, 64 threads, 128\,GB RAM).

\textbf{Provenance of the numbers.} Every figure we report for SeQLean is measured by us, under the system budgets given below. Every aggregate figure we report for a baseline is quoted from that system's own published evaluation and artifact~\citep{he2024verieql,he2024verieqlartifact,wang2024qed} rather than from our reproduction of it. We adopt this convention deliberately: each baseline was tuned and configured by its authors on precisely these corpora, so their reported coverage is the strongest and fairest statement of what the system achieves, and quoting it removes any question of our having under-configured a competitor. Where we quote a verdict on an \emph{individual} pair --- in Appendices~\ref{sec:trusted-constraints} and~\ref{sec:crossskill-appendix}, for queries that lie outside the published corpora and therefore have no published verdict --- that verdict comes from our own run under the configuration described by us, and is identified as such at the point of use.

\textbf{SeQLean.} Three mechanisms are run over the generated Lean files of Section~\ref{sec:goalgen} and reported separately rather than aggregated: \texttt{sql\_equiv}, with a budget of $10^6$ heartbeats and 300 seconds per pair; \texttt{plausible\_sql}, which searches for a counterexample within the same 300 seconds; and \texttt{sql\_equiv\_llm}, limited to five refinement rounds and 700 seconds. Each pair is compiled in a dedicated process group, so that a wall-clock kill terminates \texttt{lake} together with the \texttt{lean} process it spawns; terminating the parent alone leaves orphaned processes holding the output pipe open, which stalls the run without reporting an error. The Lean census is allocated 24 parallel workers and each baseline 16. The LLM pass over both corpora (461 files) required 1.71 hours of wall time. The deterministic passes dominate the total compute cost, as a pair that cannot be proved consumes its entire budget before failing.

\textbf{VeriEQL} accepts the two queries and the schema essentially unmodified. For the individual-pair verdicts we quote, we configure it with a database-size bound of 2 and a 40-second per-pair timeout, and take a pair as settled when it reports \texttt{EQUIV}; its corpus-level coverage is quoted from its own evaluation, per the convention above. One adaptation is required: VeriEQL provides no floating-point sort, so \texttt{FLOAT} columns are mapped to \texttt{INT} when constructing its environment. We record this explicitly because it operates in VeriEQL's favour, admitting pairs that are, strictly speaking, no longer the original problem.

\textbf{QED} does not accept SQL as input. It expects a Calcite logical plan in JSON, and the parser that produces one requires a Java runtime unavailable in our environment, so we reconstruct the plan from each query's \texttt{sqlglot} syntax tree. QED is given 60 seconds per pair with a 15-second SMT budget, and a pair is counted when it reports \texttt{PROVABLE}. Our converter is deliberately conservative: constructs outside the fragment it handles --- window functions, correlated subqueries and \texttt{EXISTS}, \texttt{VALUES}, \texttt{LATERAL}, and \texttt{LIMIT} --- are reported as unsupported rather than approximated. The resulting bias is one-directional: since QED is sound, an omission in our converter can only reduce its coverage and can never yield a spurious equivalence. QED's figures should therefore be read as a lower bound on what it achieves with its native parser. To quantify that gap, we additionally run \texttt{qed-prover} on QED's own distributed Calcite plan corpus, for which no conversion is performed.

\textbf{Evaluating the Results.} Results are joined on (corpus, pair id), so that each pair yields a single record of the outcome obtained by all three systems on an identical input. Throughout, we distinguish a pair that a system cannot encode from one it encodes but fails to prove. Conflating the two would overstate the performance of every system considered, and of ours in particular, since our dominant failure mode is an encoding failure rather than a proving failure (\textbf{UNSUPPORTED SYNTAX} in Section~\ref{sec:classification}).

\subsection{Glossary of reported outcomes, and what each census covered}
\label{sec:glossary}

Because the same words are used loosely in this literature, we fix them here. Every outcome below is a property of a \emph{pair} under a \emph{stated budget}, not a property of the queries in the abstract: a pair that times out is one our search did not finish, and says nothing about whether the two queries are equivalent.

\begin{itemize}\itemsep2pt
\item \textbf{PROVED} (equivalently \emph{Eq.}) --- a Lean proof that the two queries return the same set of rows on \emph{every} database, closed by \texttt{sql\_equiv} or by the LLM tier and rechecked by the kernel. Unbounded and not relative to any database size.
\item \textbf{REFUTED} --- a concrete database, produced by search rather than by hand, on which the two queries return different rows, together with a kernel-accepted \texttt{decide} proof of that difference. A decided negative, not a failure to prove.
\item \textbf{RESOLVED} --- PROVED \emph{or} REFUTED: the pair was decided in one direction or the other. This is the only aggregate we report as coverage, and it is never used as a synonym for PROVED.
\item \textbf{UNSUPPORTED SYNTAX} --- the pair never became a goal, because a construct in it lies outside the grammar or has no elaboration. A limit of the front end; widening the parser turns these into goals directly. It is not a proving failure and we never count it as one.
\item \textbf{ELAB-TIMEOUT} --- the budget was exhausted while turning SQL text into a goal, before any proof search began.
\item \textbf{PROVE-TIMEOUT} --- a goal existed and the tactic was \emph{interrupted still searching} when the budget ran out, leaving no verdict. Both the wall-clock cap and Lean's deterministic heartbeat limit produce this, and the latter must be told apart from UNSUPPORTED SYNTAX by its diagnostic, since it exits with the same status.
\item \textbf{UNPROVED} --- a goal existed and the tactic \emph{stopped on its own}, having run out of strategies rather than out of budget, returning neither a proof nor a counterexample. The contrast with PROVE-TIMEOUT is not one of degree: more time can resolve a PROVE-TIMEOUT and cannot resolve an UNPROVED, so this is the only category that measures proving power, and we keep it separate even where it is small.
\end{itemize}

\paragraph{Coverage of the two censuses.} Both ran to completion over all $461$ pairs, and neither was sampled or truncated. The deterministic census compiles one file per pair with \texttt{sql\_equiv} under $10^6$ heartbeats and a $300$-second wall clock. The LLM census makes a separate pass over the same $461$ files, and its own step~$0$ is a \emph{weaker} deterministic tactic than the full \texttt{sql\_equiv} of the first census --- it omits the membership route --- so its raw proved counts ($68$ on Calcite, $9$ on Literature at step~$0$, rising to $91$ and $27$ with the LLM) are not comparable with the first census and are not reported as such. What Table~\ref{tab:corpora} credits to the LLM is the marginal set: pairs it closes that the \emph{full} \texttt{sql\_equiv} does not, which is $2$ on Calcite and $1$ on Literature. Reporting the raw difference instead would double-count the pairs both tiers reach.

The refutation side of the LLM census is filtered rather than taken at face value. Of its Calcite disproofs, none passed artifact verification and all are excluded; on Literature $14$ of $16$ verified and are counted. A further $32$ Calcite and $2$ Literature candidate witnesses were rejected as spurious by the harness itself --- cases where the queries agree as real SQL and the disagreement is ours. Taken with the $15$ withdrawn refutations of Appendix~\ref{sec:status-appendix}, the pattern is consistent: a corpus of known-equivalent pairs is a test of our modelling, and on Calcite it currently finds faults in it.



\end{document}